\documentclass[fleqn,10pt]{wlscirep}
\usepackage[utf8]{inputenc}
\usepackage[T1]{fontenc}
\usepackage{array}
\usepackage{todonotes}
\usepackage{textgreek}
\usepackage{lscape}
\usepackage{graphicx}
\usepackage{lipsum}
\usepackage{ccaption}
\usepackage{soul}
\usepackage{float}
\usepackage{amsmath}
\usepackage{graphicx}
\usepackage{subcaption}
\usepackage{algorithm} 
\usepackage{algpseudocode} 
\usepackage{makecell}
\usepackage{placeins}

\newcolumntype{M}[1]{>{\centering\arraybackslash}m{#1}}

\newcommand{\beginsupplement}{%
        \setcounter{table}{0}
        \renewcommand{\thetable}{\S\arabic{table}}%
        \setcounter{figure}{0}
        \renewcommand{\thefigure}{S\arabic{figure}}%
     }

\title{\LARGE Brain-wide interpolation and conditioning of gene expression in the human brain using Implicit Neural Representations 
\vspace{1em}}

\author[1,5]{\large Xizheng Yu}
\author[2]{\large Justin Torok}
\author[3]{\large Sneha Pandya}
\author[1]{\large Sourav Pal}
\author[4,1*]{\large Vikas Singh}
\author[2*]{\large Ashish Raj}
\affil[1]{Department of Computer Sciences, University of Wisconsin - Madison, Madison, WI, USA}
\affil[2]{Department of Radiology, University of California, San Francisco, CA, USA}
\affil[3]{Department of Radiology, Weill-Cornell Medical College, New York, USA}
\affil[4]{Department of Biostatistics, University of Wisconsin-Madison, Madison, WI, USA}
\affil[5]{Department of Computer Science, Brown University, Providence, RI, USA}
\affil[*]{Address correspondence to ashish.raj@ucsf.edu and vsingh@biostat.wisc.edu}

\begin{abstract}

In this paper, we study the efficacy and utility of recent advances in non-local, non-linear image interpolation and extrapolation algorithms, specifically, ideas based on Implicit Neural Representations (INR), as a tool for analysis of spatial transcriptomics data. We seek to utilize the microarray gene expression data sparsely sampled in the healthy human brain, and produce fully resolved spatial maps of any given gene across the whole brain at a voxel-level resolution. To do so, we first obtained the 100 top AD risk genes, whose baseline spatial transcriptional profiles were obtained from the Allen Human Brain Atlas (AHBA). We adapted Implicit Neural Representation models so that the pipeline can produce robust voxel-resolution quantitative maps of all genes. We present a variety of experiments using interpolations obtained from \texttt{abagen} as a baseline/reference.
 
\end{abstract}

\begin{document}

\flushbottom
\maketitle
\thispagestyle{empty}

\section{Introduction}

The spatial, anatomic and connectivity architecture of the brain is fundamentally related to its function. Spatial transcriptomics (ST) -- the measurement and analysis of spatially-varying transcriptional signatures, e.g. via RNA sequencing of microarrays or \textit{in situ} hybridization -- can link brain cell types defined by molecular composition to their morphology, activity, and connectivity \cite{lein_promise_2017}. ST has emerged as a powerful tool for investigating how structural and molecular organization governs or mediates brain function, development and disease within their native tissue context, particularly in complex organs like the brain \cite{lein_promise_2017, park_spatial_2023, jung_spatial_2023}. This technology enables the creation of molecular maps that can redefine neuroanatomy by providing unbiased, high-resolution spatial information about cell types, connectivity, and function \cite{ortiz_spatial_2021}. Recent ST studies have also led to unprecedented insights into neurological disorders such as glioblastoma \cite{jung_spatial_2023, shireman_spatial_2023}. In glioblastoma research, ST has revealed how tumors integrate into and remodel the surrounding brain tissue, offering potential avenues for therapeutic breakthroughs \cite{shireman_spatial_2023}. ST is a powerful tool in understanding dementias \cite{sepulcre_neurogenetic_2018, anand2024selective, shafiei2023network}. In parallel, our understanding of the cell types that make up the brain is rapidly accelerating, driven in particular by recent advances in single-cell ST that enable the spatially-resolved molecular profiling of the human brain in both healthy \cite{Hodge2019, Bakken2021} and disease conditions \cite{fu2018selective, leng2021molecular, gabitto2024integrated, kamath2022single}.


With the improving spatial resolution and coverage of ST, it is becoming possible to probe neuroscientific questions at finer and finer spatial scales. However, it is still not possible to sample all spatial sites densely in any organ, and especially so in the brain. In particular, single-cell sequencing is limited by the number of sampling sites due to technical and cost issues. The well-established AHBA~\cite{Hawrylycz_atlas_human_brain_transcriptome_2012}, which sampled 926 discrete brain locations, provides far greater spatial coverage than historically possible, but it is still insufficient beyond coarse region-level analyses.
In addition to experimentally increasing sampling density, many laboratories are therefore exploring algorithmic means of improving the spatial sampling density and coverage of gene expression data, using various interpolation techniques. 


Current interpolation methods for mapping brain gene expression have primarily relied on traditional techniques such as nearest-neighbors interpolation \cite{Whitaker2016-af} and linear interpolation \cite{Burt2018}. This is the approach taken by a widely-adopted tool called \texttt{abagen} \cite{markello2021standardizing}. However, these local interpolation methods often fail to capture the complex spatial relationships and variations in gene expression across different brain regions. Nearest-neighbors interpolation can oversimplify the data by assigning the same value to adjacent areas. Linear interpolation, while more refined, still assumes a smooth change between data points, which does not adequately represent the nonlinear patterns often present in gene expression data. Gaussian process regression (GPR) was proposed to infer voxel-wise maps of AHBA genes, but did not perform well for genes with low spatial variation (relative structured variability) given strength of spatial dependence \cite{gregor2018gpr}. GPR relies on the assumption that proximate samples have similar values and requires predefined kernel functions to model spatial relationships \cite{gregor2018gpr}. This can limit the model's flexibility, especially when dealing with complex, high-dimensional data like 3D brain gene maps. 
Given these limitations of interpolation from limited samples, the prospect of super-resolution -- recovery of signals from experimental data below the Nyquist sampling rate -- remains out of reach for ST in the human CNS. 
While compressive sensing is effective in recovering the unobserved signal from a set of extremely sparse (in some basis) measurements, and works well in many other domains \cite{donoho2006, bora2017compresse}, its applicability to ST is unclear (is not obvious whether there is an appropriate sparse basis to deploy CS).

\begin{figure}[!t]
    \centering
    \includegraphics[width=\textwidth]{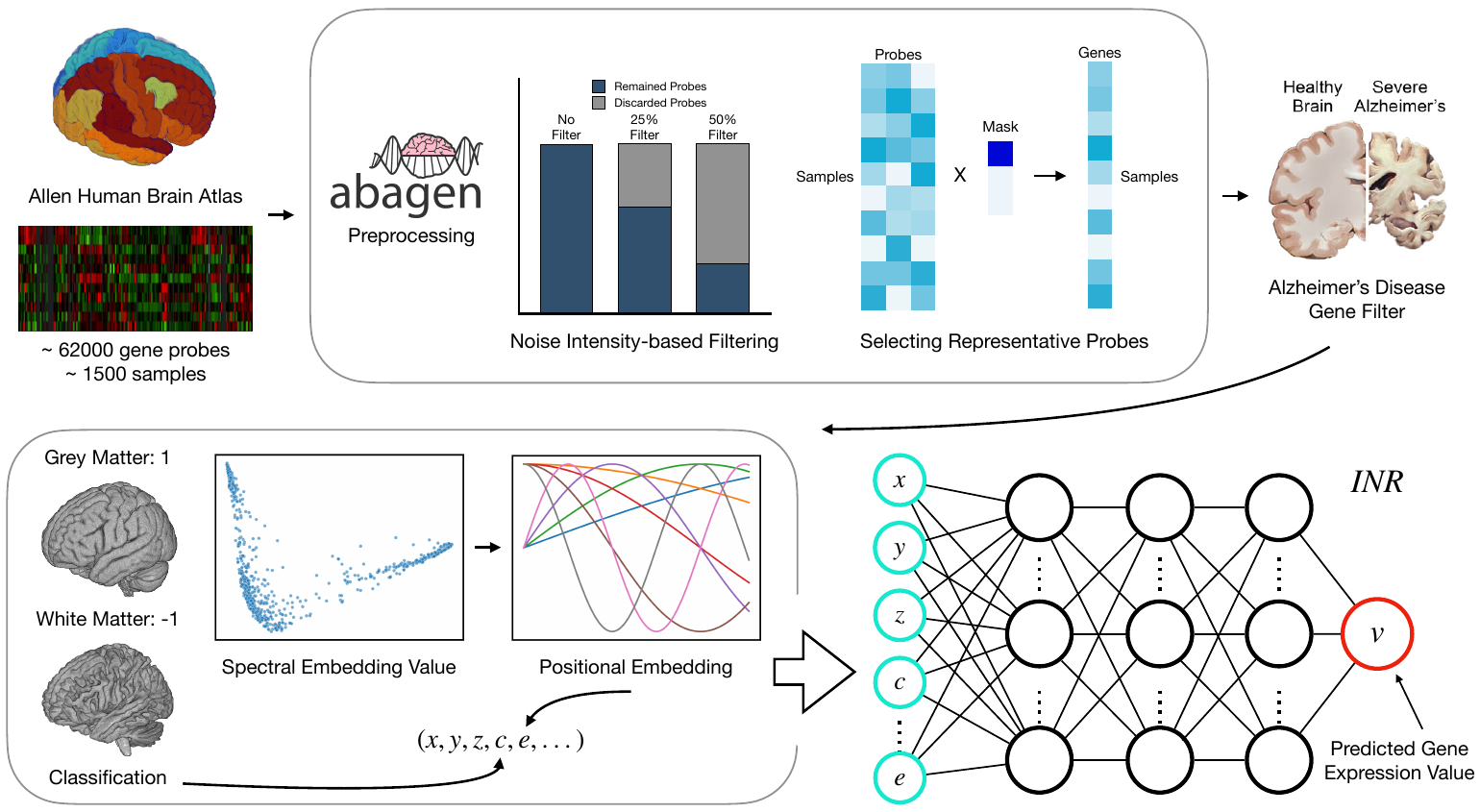}
    \caption{Flow chart of the INR interpolation pipeline: Starting from the Allen Human Brain Atlas (left), which contains approximately 62,000 gene probes from 1,500 tissue samples, the data undergoes preprocessing through the \texttt{abagen} toolkit (Noise Intensity-based Filtering and Representative Probes Selection and so on). An Alzheimer's Disease filter is then applied to focus on AD-relevant genes. Inputs to the Implicit Neural Representation (INR) network (left bottom): spatial coordinates ($x$, $y$, $z$), tissue classification ($c$, where grey matter = 1, white matter = -1), and gene spectral embedding dimension ($e$), which is enhanced through positional embedding to capture multi-scale relationships. Input vectors $(x, y, z, c, e, ...)$ feed to the INR network and the network outputs predicted gene expression values ($v$) at any queried coordinate location, enabling continuous voxel-wise interpolation across the entire brain volume. The detailed algorithm underlying this workflow is provided in Algorithm \ref{algo:flow} of the Supplementary Information.}
    \label{fig:flow_chart}
\end{figure} \noindent

\subsection*{Implicit Neural Representation (INR) models for gene interpolation and super-resolution}

Overcoming the limitations of existing interpolation techniques requires an approach that moves away from assumptions about the underlying spatial structure of the data, explicit forms of statistical dependencies or reliance on predefined basis functions. To achieve these goals, we propose the use of a framework designed to learn its own data-driven basis via a combination of activation functions and weights. The basis will offer the ability to capture complex and non-linear patterns over a long receptive field through universal function approximation capabilities available in deep learning architectures \cite{}. Our formulation, based on developments in Implicit Neural Representation (INR), will enable feature sharing across spatial locations to achieve an implicit regularization useful to fit variations across regions and spatial scales. INRs do not suffer from the inherent shortcomings of local filtering/interpolation \cite{Markello_abagen_2021} or Gaussian Process Regression (GPR) \cite{gregor2018gpr} because the INR model learns to associate 
positions (measurement locations) directly 
to the measurement, 
and once trained, the model (which parameterizes a continuous function) can be probed to provide interpolated values of the function (i.e. gene expression measurements) at other brain locations that were not sampled. 
Our goal 
is to develop a non-local, non-linear INR-based technique for super-resolution and brain-wide mapping of spatial transcriptomics, and assess its efficacy on the best-available public ST datasets in the human CNS, specifically, the Allen Human Brain Atlas (AHBA) from the Allen Institute \cite{Hawrylycz_atlas_human_brain_transcriptome_2012}. We use microarray gene expression data as an input, sparsely sampled in the healthy human brain, and produce fully resolved spatial maps of any given gene across the whole brain at a voxel-level resolution. 

In addition to the adaptation of INR for the ST task, we must also use other strategies to overcome the extremely sparse sampling available in ABA. Specifically, we propose \emph{gene profile axis as an auxiliary signal} to overcome sampling limitations. To enrich the feature space that our INR can be trained on, we will use an ordinal index that is designed to aid the interpolation process: the distance along the gene profile axis of the similarity between genes. The rationale behind this augmentation is that an individual gene may not have sufficient information to successfully overcome the undersampling inherent in the gene expression data but by exploiting the fact that genes close to each other in the gene expression axis should also be close in the spatial domain. 
This yields meaningful guidance to the INR model achieved via a spectral embedding of expression data across a range of genes involved in a related biological processes. 

{\bf Paper organization.} We first review the INR methodology, then propose our implementation and finally assess its performance on ABA data. We then compare our maps to the region-level maps obtained previously by a well-established and widely used interpolation tool called \texttt{abagen} \cite{markello2021standardizing}. Since there is no ground truth available for ABA gene expression we instead rely on indirect assessments to demonstrate the concept validity of the proposed technique. For this purpose, we investigated risk genes that are known to be involved in Alzheimer's disease (AD). We show that not only are our maps strongly correlated with the \texttt{abagen} maps, they also possess very strong and clear spatial association with AD pathologic hallmark, specifically, tau-PET uptake. We note however that the technique is completely agnostic to the specific choice of genes and can be easily extended to other disease contexts.

\section{Materials and Methods}

\begin{figure}[!t]
    \centering
    \includegraphics[width=1\textwidth]{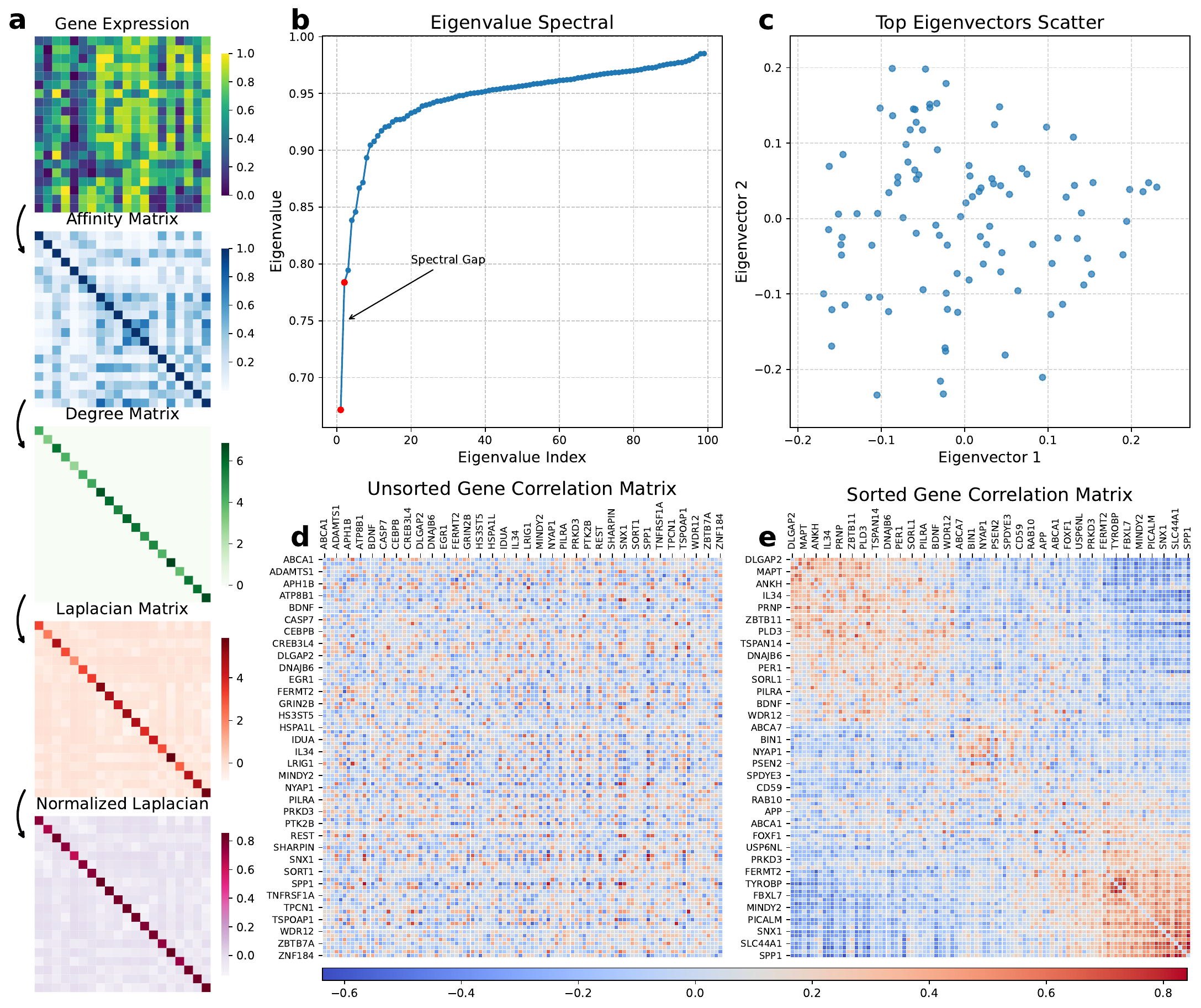}
    \caption{\textbf{a.} Visualization of the spectral embedding pipeline: starting with raw gene expression data, calculating the affinity matrix representing gene-to-gene similarity, constructing the degree matrix, deriving the Laplacian matrix, and finally obtaining the normalized Laplacian used for embedding. \textbf{b.} Eigenvalue spectrum showing a clear spectral gap between the top eigenvalues and the rest. \textbf{c.} Scatter plot of gene positions using the top two eigenvectors, demonstrating the distributed representation of genes in the embedding space. \textbf{d.} Unsorted gene correlation matrix. \textbf{e.} Sorted gene correlation matrix after applying spectral embedding, revealing clear clusters of functionally related genes. The sorted matrix demonstrates how the spectral embedding effectively captures the functional relationships between genes, with distinct blocks of positively correlated (red) and negatively correlated (blue) genes appearing after sorting according to the spectral coordinates. Graphs are generated with donor 10021's data from AHBA.}
    \label{fig:spectral}
\end{figure}

\subsection{Gene expression data}

\textbf{Regional AD gene expression}. 
We identified prominent AD genes from the list of AD loci and genes compiled by Alzheimer's Disease Sequencing Project (ADSP) Gene Verification Committee as well as referenced from various studies \cite{Karch_AD_genetics_2014, Ayoub_2023, Seto_genes_AD_2021, Andrews_GWAS_risk_2020, Dumitrescu_resilience_AD_2020}. A total of $100$ genes were shortlisted (Supplementary Table\ref{tab:gene_names_100_se} and Table\ref{tab:gene_names}). For each gene, we obtained healthy baseline expression levels from AHBA using the procedure described below. For both the final training gene expression matrix and interpolation gene expression matrix, we chose 100 AD genes giving us a gene expression table.

\noindent \textbf{Gene expression data preprocessing}. The publicly available Allen Human Brain Atlas (AHBA) \cite{Hawrylycz_atlas_human_brain_transcriptome_2012} includes 926 brain regions, with each region having microarray expression levels from a set of 58,692 probes that correspond to 21,245 distinct genes. Expression data for each of the 926 regions from AHBA were mapped to the 86 regions of the Desikan-Killiany atlas. All samples for all probes within the same region were averaged and then normalized for each gene to produce a single expression value quantified as a $z$-score. White matter tracts were excluded from the analysis. 
Regional microarray expression data were obtained from 6 post-mortem brains (2 with bilateral hemisphere samples and 4 with left hemisphere samples only; one female, ages 24.0--57.0) provided by the publicly available Allen Human Brain Atlas \cite{Hawrylycz_atlas_human_brain_transcriptome_2012} (AHBA, https://human.brain-map.org). To ensure result stability and consistency across hemispheres, the main analyses were conducted using the two male donors (ages 24.0--39.0) with bilateral hemisphere samples.
We used the \texttt{abagen} toolkit \cite{Markello_abagen_2021} (version 0.1.4; https://github.com/rmarkello/abagen) to preprocess this dataset, and the steps in our pipeline are described next. 

Samples with mismatched MNI coordinates that did not align with the specified ontology were excluded. The MNI coordinates of tissue samples were updated to those generated via non-linear registration using the Advanced Normalization Tools (ANTs; https://github.com/chrisfilo/alleninf). Samples were assigned to brain regions in the provided atlas if their MNI coordinates were within 2 mm of a given parcel. Sample-to-region matching was constrained by hemisphere and gross structural divisions to reduce the potential for misassignment (i.e., cortex, subcortex/brainstem, and cerebellum, such that e.g., a sample in the left cortex could only be assigned to an atlas parcel in the left cortex \cite{Arnatkeviciute_gene_exp_neuroim_2019}).
Then, we averaged the microarray expression data across probes for each donor. Microarray probes were first reannotated using data provided in published works  \cite{Arnatkeviciute_gene_exp_neuroim_2019}, after which probes not matched to a valid Entrez ID were discarded. Probes were filtered based on their expression intensity relative to background noise \cite{Pandya_seed_spread_ALS_2022}, such that probes with intensity less than the background in $\geq$50\% of samples across donors were discarded, yielding 30,534 probes in all. When multiple probes indexed the expression of the same gene, we selected the probe with the most consistent pattern of regional variation across donors \cite{hawrylycz2015canonical}. This consistency was determined by measuring the average pairwise correlation of regional expression profiles between donors, allowing us to identify the most representative probe for each gene. Regions in this analysis correspond to the structural designations provided in the ontology from the Allen Human Brain Atlas (AHBA).
Next, we structured our data into normalized gene expression values \(v\). Inter-subject variation was addressed by normalizing tissue sample expression values across genes using a robust sigmoid function: $x_{norm} = 1/({1 + \exp(-\frac{(x-\langle x \rangle)}{\text{IQR}_{x}})})$ where $\langle x \rangle$ is the median and $\text{IQR}_{x}$ is the normalized interquartile range of the expression of a single tissue sample across genes. Normalized expression values were then rescaled to the unit interval using min-max normalization. Gene expression values were normalized across tissue samples using an identical procedure. Normalization was performed separately for samples in distinct structural classes (i.e., cortex, subcortex/brainstem, cerebellum) and tissue samples not matched to a brain region were discarded after normalization.

\noindent \textbf{Voxel-wise Gene Expression}.
The \texttt{abagen} process has been described in detail in the literature \cite{Markello_abagen_2021, Pandya_seed_spread_ALS_2022}. Briefly, microarray probes were reannotated using the data made available \cite{Arnatkeviciute_gene_exp_neuroim_2019}. As noted above, filtering the probes based on their expression intensity relative to background noise\cite{Pandya_seed_spread_ALS_2022} yielded $30534$ probes. To obtain voxel-wise data for training, we set \texttt{region\_agg=None} and \texttt{return\_donors=True}, and saved the resulting data (as a CSV file). This setting skips region aggregation across probes and allows us to obtain probe-level, noise-filtered expression data for each donor prior to any regional aggregation. Only the first two donors included in the AHBA had tissue samples taken from the right hemisphere, for the four donors with data available for only the left hemisphere, we used \texttt{lr\_mirror='leftright'} to mirror existing tissue samples across the left-right hemisphere boundary \cite{romero2018structural}. This provided us with raw voxel-level gene expression data with corresponding spatial coordinates for model training, with variable number of rows, corresponding to voxel locations before brain region aggregation, and 15,329 columns, corresponding to the retained genes. 

\noindent \textbf{Interpolation of Gene Expression},
For comparison with our INR results, we also generated a donor-wise interpolated result. For this interpolation process, we removed the aforementioned parameters and added \texttt{missing='interpolate'} to interpolate expression values in the empty regions by assigning every node in the region the expression of the nearest sample and taking a weighted (inverse distance) average \cite{Markello_abagen_2021}. If a brain region was not assigned a tissue sample based on the above procedure, every voxel in the region was mapped to the nearest tissue sample from the donor in order to generate a dense, interpolated expression map. The average of these expression values was calculated across all voxels in the region, weighted by the distance between each voxel and the sample mapped to it, to determine an estimate of the parcellated expression values for the missing region. Samples assigned to the same brain region were averaged separately for each donor \texttt{return\_donors=True}, yielding 6 regional expression matrices with 86 rows, corresponding to brain regions, and 15,329 columns, corresponding to the retained genes.

\subsection{Alzheimer-associated tau uptake from ADNI}

AD-associated tau data used in this work were obtained from the Alzheimer’s Disease Neuroimaging Initiative (ADNI3) database (\url {adni.loni.usc.edu}), constituting a sample size of 196 participants with AD and those in early and late stages of mild cognitive impairment (EMCI and LMCI). Of those, 102 participants were labeled as EMCI (64 men/38 women, mean age 77.73 $\pm$ 7.17), 47 participants as LMCI (29 men/18 women, mean age 75.94 $\pm$ 7.56), and 47 participants as AD (25 men/22 women, mean age 78.43 $\pm$ 9.17). High-resolution T1-weighted sagittal brain MR and AV1451-PET ('tau-PET') imaging data were obtained from ADNI. 
\\
\textbf{Brain regional parcellation and tau co-registration}.
The MRI data in our analysis was based on a 3D MPRAGE sequence with the following specific parameters: an 8-channel coil, TR (Repetition Time) of 400ms, minimum full TE (Echo Time), 11 degrees flip angle, slice thickness of 1.2mm, resolution of $256 \times 256$ mm, and a FOV (Field of View) of 26cm. Standardized image processing pipeline was followed: T1 images underwent normalization to the Montreal Neurological Institute (MNI) space and segmentation using SPM8's unified coregistration and segmentation scheme (\url {https://www.fil.ion.ucl.ac.uk/spm/}). Grey matter (GM) was divided into 86 regions of interest (ROIs) based on the Desikan-Killiany (DK) atlas \cite{Desikan_2006}. Tau-PET images were normalized using cerebellar uptake, and resliced to ensure uniform voxel resolution. To establish a reference point, PET images from healthy controls were utilized to generate an average image, which was then normalized to the MNI space using SPM8's linear and non-linear transformations. This transformation was consistently applied to all individual PET images, aligning them with the same resolution as the 86-region DK atlas, resulting in a $86\times 1$ regional tau vector per subject. 
Finally, all AD and MCI participants' regional tau vectors were averaged to produce a single {\emph AD-associated tau regional vector} $\mathbf{\tau} \in \mathcal{R}^N$. Group averaging here is appropriate because we wish to compare tau against regional gene expression, which is not available for individual subjects. 

\section{INR based Interpolation}

\subsection{Definitions}

\noindent \textbf{Variables and Notation.} Let $\mathbf{x} = (x, y, z)$ represent a 3D coordinate in MNI space within the brain, where $x$, $y$, and $z$ are spatial coordinates. Let $g$ represent a specific gene, and $v$ represent the normalized gene expression level at a given location for a specific gene. Our goal is to learn a function, $\Phi$, that maps this coordinate and gene information to the normalized gene expression level as $\Phi(\mathbf{x}, g) = v$. We denote the parameterized version of this function as $\Phi_\theta$, where $\theta$ represents the parameters (weights and biases) of the neural network used to implement the function.

\noindent \textbf{Spectral Embedding}. 
Spectral embedding is a dimensionality reduction technique that uses the eigendecomposition (spectral) of matrices derived from the data, typically the graph Laplacian matrix. Rather than seeking to preserve pairwise distances (prominent in methods like multi-dimensional scaling), spectral embedding specifically preserves the local neighborhood structure and connectivity patterns in the data.
We employ spectral embedding with gene expression data and construct a series of matrices to capture relationships between genes (see Figure \ref{fig:spectral}a) as described below. 

First, we set up an affinity matrix \(A\) where each element \(A_{ij}\) represents the similarity between genes \(i\) and \(j\). Then, a degree matrix \(D\) is derived as a diagonal matrix where \(D_{ii} = \sum_j A_{ij}\), representing the total connection strength of each gene to all others. The Laplacian matrix \(L\) is then calculated as \(L = D - A\), effectively encoding both the connection patterns and strengths between genes. For numerical stability, we use the normalized Laplacian \cite{scikit-learn} defined as 
\( L_{\text{norm}} = I - D^{-1/2} A D^{-1/2} \).
The embedding will be based on the eigenvalues and eigenvectors of a Laplacian matrix (see Figure \ref{fig:spectral}b,c): derived from the data. Formally, spectral embedding is written as:
$$ \mathbf{L} \mathbf{v}_i = \lambda_i \mathbf{v}_i, $$
where \( \mathbf{L} \) is the Laplacian matrix, \( \lambda_i \) are the eigenvalues, and \( \mathbf{v}_i \) are the corresponding eigenvectors. The eigenvectors corresponding to the smallest non-zero, or non-trivial eigenvalues are used as the new coordinates in the embedded space. The presence of a spectral gap, as shown in Figure (see Figure \ref{fig:spectral}b), indicates natural clusters within the data and helps determine the optimal dimensionality of the embedding space. Using this procedure, we can apply  dimensionality reduction on our gene representations and use the embedding obtained to provide an ordinal value for the genes.

We apply spectral embedding \cite{ng2001spectral} to the top 100 genes (Figure \ref{fig:spectral}), sorted by the first eigenvector. Specifically, the gene embedding $e_g$ is defined as $e_g$ = $\mathbf{v_1}(g)$, where $\mathbf{v_1}$ is the first eigenvector (corresponding to the smallest non-zero eigenvalue) and $g$ is the index of gene $g$ in the eigenvector. We normalize $e_g$ using its mean and standard deviation to scale its values to the interval $[0,1]$. The finalized training data thus consists of inputs in the form of $(x, y, z, c, e_g)$ and outputs $v$, where $e_g$ is the spectral embedding representation of gene $g$. Our function $\Phi_\theta$ can therefore be more precisely expressed as $\Phi_\theta(x, y, z, c, e_g) = v$.

\noindent \textbf{Model Inputs}. To give our model additional anatomical context, we use an additional anatomical dimension $c$, which identifies whether a coordinate belongs to grey matter ($c=1$), white matter ($c=-1$), or neither ($c=0$). Further, as described above, to capture the functional continuum between genes, we use an extra dimension $e_g$, which represents the gene embeddings. 

\subsection{INR Model for Gene Expression Interpolation}

{\bf Background of INRs}. Implicit Neural Representations (INRs) offer a powerful framework to represent complex, continuous functions using neural networks. The INR model was initially introduced to represent complex signals like 3D shapes, images, music as continuous functions using neural networks \cite{sitzmann2020implicit}.
While neural networks such as multilayer perceptrons (MLPs) are often 
used for learning representations for data, 
in INRs, one feeds discrete data such as coordinates within an image, 
to the MLP. 
The MLP then estimates the parameters needed to non-linearly transform these coordinates into a continuous representation, e.g., the RGB color 
observed at that pixel coordinate for an image. 

A specific example of INR 
is the SIREN model \cite{sitzmann2020implicit}, 
which passes the input coordinates 
through sinusoidal activations, 
to effectively capture the mapping between the coordinates and 
the observed color at that specific pixel. What this process accomplishes 
is that the MLP, once its parameters are 
estimated, serves as a continuous representation of the image its parameters were estimated for; in other words, 
any arbitrary coordinate can now be queried, and the output color value 
for the pixel will be the result of 
an interpolation based on the INR's internal representation.
From a practical standpoint, INRs can take  advantage of hardware accelerators like GPUs, making them more practical for handling large datasets and high-resolution predictions across the brain. In contrast, GPR is less suitable for such tasks unless specific code optimizations 
are done.

{\bf INR modeling for gene expression interpolation}. Unlike alternative methods that discretize space (e.g., into voxels), INRs define a function {\em implicitly}. For gene expression, this means that after the model estimation is done, we can query the expression level of a specific gene at any continuous 3D coordinate within the brain, not just at a limited set of discretized locations. In the original result introducing the idea \cite{sitzmann2020implicit}, the general INR formulation takes the form, $F \left( \mathbf{x}, \Phi, \nabla_{\mathbf{x}} \Phi, \nabla^2_{\mathbf{x}} \Phi, \ldots \right) = 0, \quad \Phi : \mathbf{x} \mapsto \Phi(\mathbf{x}).$ Here, $F$ is a function that relates $\mathbf{x}$, $\Phi$, and its derivatives. The expression $F(\cdot) = 0$ defines the {\em implicit} relationship. This general form is powerful because it can represent a wide range of functions. For the specific INR model choice, we use Sinusoidal Representation Networks (SIREN) \cite{sitzmann2020implicit}, which is leverages periodic functions as activation functions. The SIREN activation is defined as: $ \phi_i (\mathbf{x}_i) = \sin (\mathbf{W}_i \mathbf{x}_i + \mathbf{b}_i) $ where $\mathbf{W}_i$ and $\mathbf{b}_i$ are the weights and biases of the network. SIREN offers several advantages,  
e.g., the periodic nature of sine functions allows it to effectively capture high-frequency details and patterns in the signal.

\begin{figure}[!t]
    \centering
    \includegraphics[width=1\textwidth]{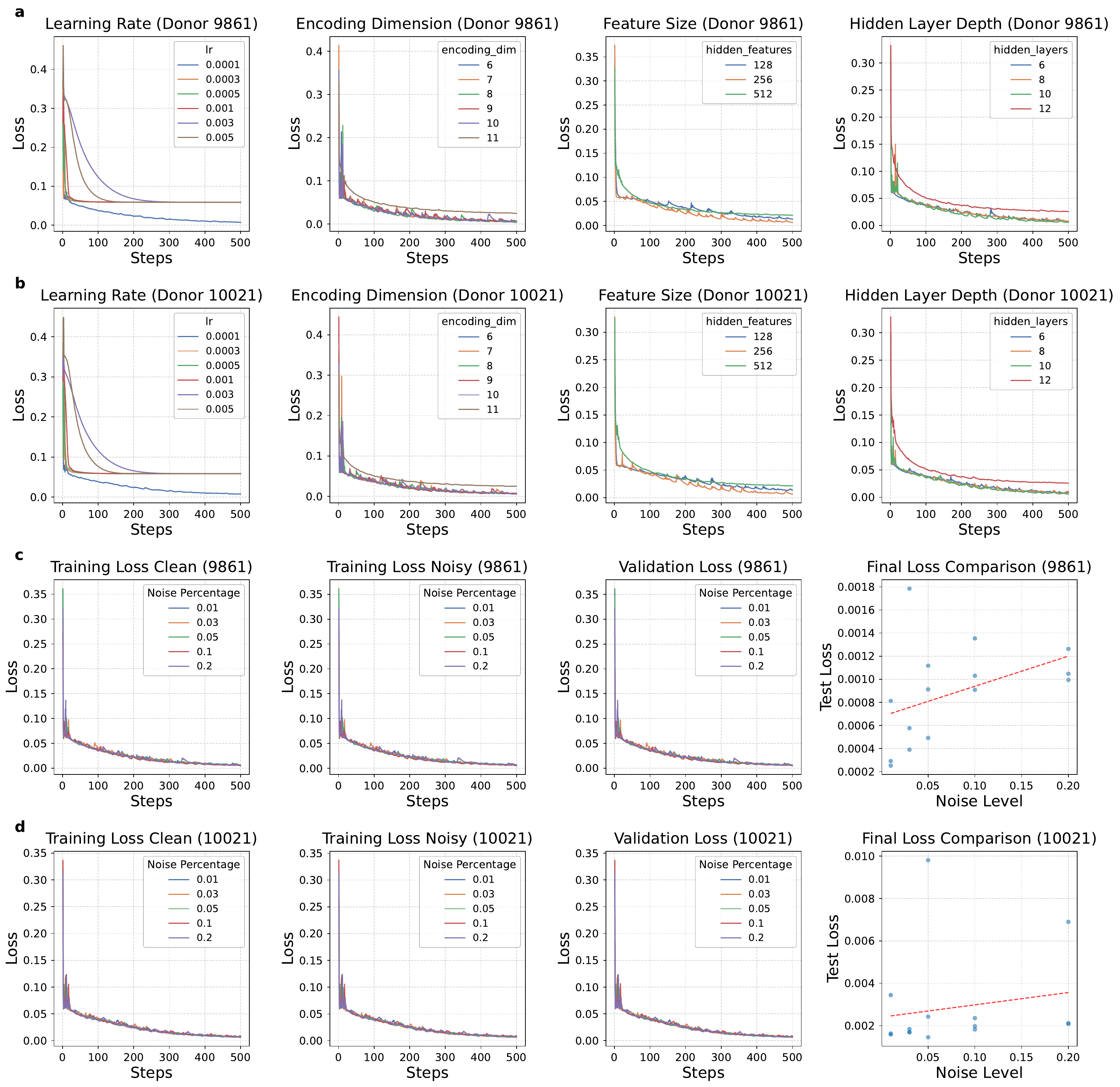}
    \caption{Comprehensive ablation study of model parameters and robustness analysis across two donors (9861 and 10021). \textbf{a} and \textbf{b} present parameter optimization studies for each donor separately, examining learning rates, encodding dimension, feature size of network, and hidden layer depther. We take loss values of first 500 steps to show general pattern and performance from the experiment. In general, learning rate at $1 \times 10^{-4}$ brings stable performance and convergence, bigger encoding dimention, feature size, deeper network all bring better convergence. \textbf{c} and \textbf{d} evaluate model robustness through added noise experiments, the first column we bring the training loss on clean data, then second column is training loss with varying noise level, then third column is validation loss on clean data under difference noise levels, and final loss comparation plotting noise level versus loss values, with a trend line. The results show remarkable consistency in training dynamics across both donors, with the model maintaining stable performance even under increasing noise levels. The linear trend in the final loss comparison suggests a predictable degradation in performance as noise increases, though the impact remains relatively modest, demonstrating the model's robustness to perturbations in the input data.}
    \label{fig:ablation}
\end{figure}

For gene expression interpolation, we implement the INR as a function $\Phi_\theta$ that takes the normalized spatial coordinates $(x, y, z)$, the anatomical label $c$, and the spectral embedding $e_g$ of gene $g$ as inputs. To improve the INR model’s capacity to learn lower frequency variations, we incorporate positional encoding into the spectral embedding during the training process. We define $\gamma(e_g)$ as the positional encoding function that maps this scalar input to a higher-dimensional vector. The positional encoding, adapted from the Neural Radiance Fields (NeRF) approach \cite{mildenhall2020nerf}, is defined as: $ \gamma(e_g) = \left(\sin(2^{0} \pi e_g), \cos(2^{0} \pi e_g), \ldots, \sin(2^{N-1} \pi e_g), \cos(2^{N-1} \pi e_g)\right) $ where $N$ is a hyperparameter representing the number of frequency components, resulting in a vector of length $2N$. This vector $\gamma(e_g)$ replaces $e_g$ as the new input into the function $\Phi_\theta$. We selected $N =10$ based on ablation experiments on our data (Fig. \ref{fig:ablation}) to ensure that each component captures the input at a different frequency scale, allowing the network to more easily represent functions across multiple frequency bands. The encoding scheme transforms input coordinates into higher dimensions which allows the network to capture both details and broader patterns in the data by decomposing the signal into multiple frequency components.

{\bf Relevant implementation details}. Our implementation is based on PyTorch \cite{paszke2017automatic}. Specifically, our model is given as: $\Phi_\theta: (x, y, z, c, \gamma(e_g)) \mapsto v$ where $v$ is the predicted normalized gene expression value at that location. The function $\Phi_\theta$ is implemented as a SIREN network with parameters $\theta$. We want $\Phi_\theta$ to approximate the observed gene expression data, which can be formulated as finding $\Phi_\theta$ such that:
$F((x, y, z, c, e_g), \Phi_\theta) = \Phi_\theta(x, y, z, c, \gamma(e_g)) - v \approx 0$.
This formulation simply states that the output of our INR $\Phi_\theta$ should equal the observed gene expression value $v$. In practice, we minimize a loss function that penalizes this difference across all training examples. Our INR architecture consists of 12 SIREN layers with 512 units each. It processes the input vector $(x, y, z, c, \gamma(e_g))$ through a series of SIREN layers. After processing the initial input $h_0 = (x, y, z, c, \gamma(e_g))$, each hidden layer applies a linear transformation followed by a sine activation function: $h_{i+1} = \sin(W_i h_i + b_i) \quad \text{for } i = 0, \ldots, L-2$, where $L=12$ is the total number of layers, $W_i$ and $b_i$ are the weights and biases of the $i$-th layer, and $h_i$ is the activation at the $i$-th layer. The final layer produces the predicted gene expression value $v = W_{L-1} h_{L-1} + b_{L-1}$. The output scalar value $v$ representing the predicted gene expression level. The model is trained using a learning rate of $1 \times 10^{-4}$ over 5000 steps and takes the best model (smallest loss value). This configuration was chosen to optimize the loss function.

We know that the gene expression data is highly undersampled and violates the Nyquist-Shannon sampling rate. So,  even provably good reconstruction is not possible in this regime. Instead, we aim to find a plausible/regularized continuous representation that is consistent with the available data at the measured locations. The INR, parameterized by a neural network, provides this regularization, and the specific choice of sinusoidal nonlinearities 
%
favorably alters the 
the types of functions it can represent. Layers of sinusoidal activation functions, unlike, for example, ReLUs, do not have the capacity to represent sharp discontinuities. Each sinusoidal layer has a limited bandwidth. While the composition of multiple sinusoidal layers can create more complex functions, the overall capacity of the network remains constrained by the number of layers and the magnitude of the weights. This bandwidth limitation serves as a good regularizer and helps avoid overfitting to the sparse data by creating spurious high-frequency oscillations. In other words, the network is forced to learn a smoother, lower-bandwidth representation that is consistent with the data. 

\textbf{Model Inference}. For inference, we systematically sample every possible voxel coordinate within the atlas and transform these into MNI coordinates. Coordinates corresponding to empty spaces, indicated by a value of $0$, are excluded from further analysis. For each retained coordinate $(x, y, z)$, we set $e_g$ to the spectral embedding of the gene of interest, enabling specific gene interpolation. Additionally, the anatomical label $c$ is assigned values of $\{1,-1\}$ depending on whether the voxel is located in grey or white matter, respectively, as defined by the Desikan-Killiany atlas \cite{Desikan_2006}. We then compute $\gamma(e_g)$ and feed the complete vector $(x, y, z, c, \gamma(e_g))$ into the trained model $\Phi_\theta$ to obtain the predicted gene expression value at that coordinate location. This setup facilitates coordinate-wise detailed interpolation, enabling precise predictions at the voxel level. Following this detailed interpolation, we compute an average for each gene similar to the process used by \texttt{abagen} \cite{Markello_abagen_2021}, thereby providing a region-wise interpolated result. This method ensures that our model outputs are both locally accurate and globally representative.

\begin{figure*}[!t]
    \centering
    \includegraphics[width=\textwidth]{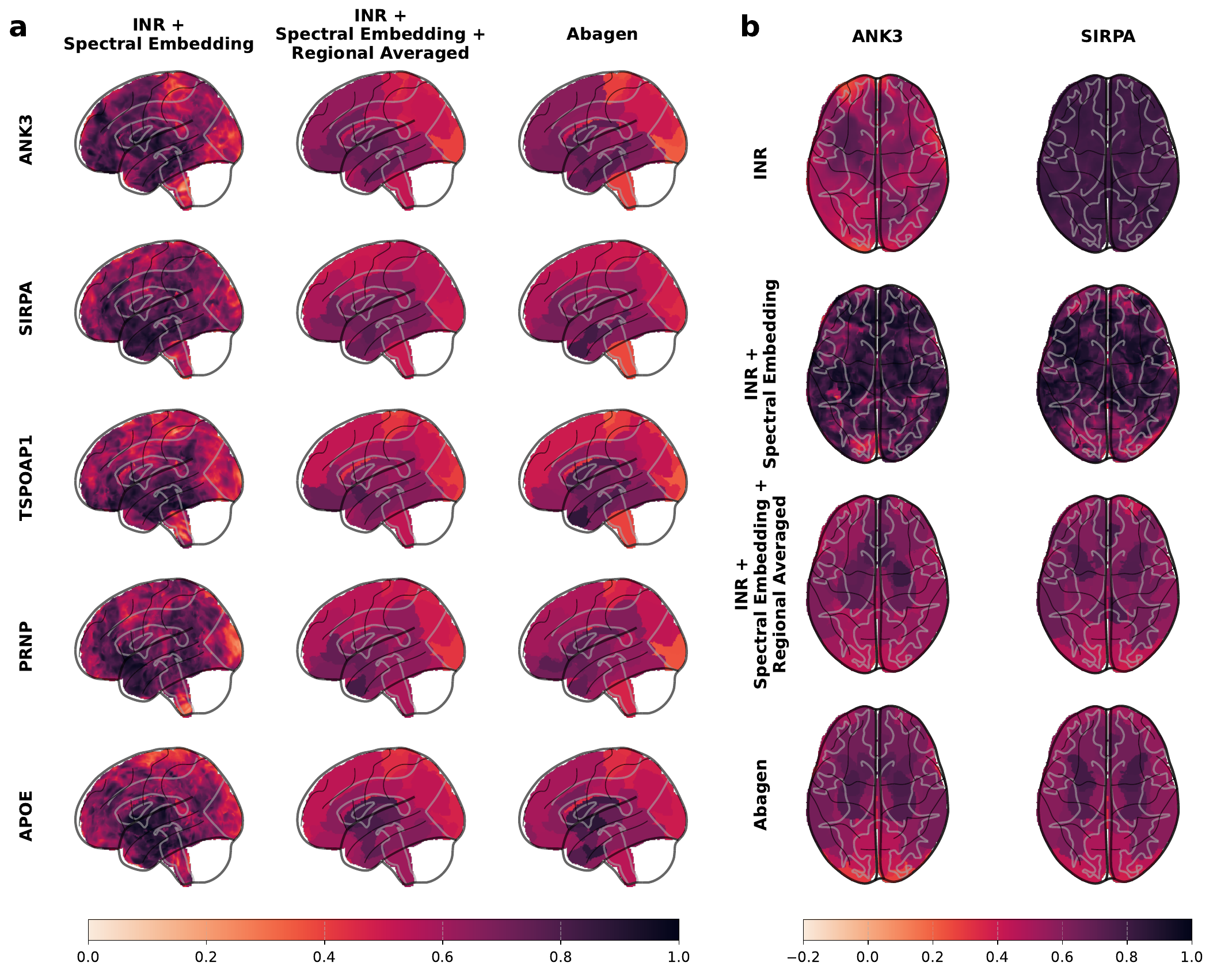}
    \caption{\textbf{a}: Sagittal view visualizations of gene expression patterns in the human brain using three different interpolation methods (INR + Spectral Embedding, INR + Spectral Embedding + Regional Averaged, and Abagen) for five tau-associated genes (ANK3, SIRPA, TSPOAP1, PRNP, and APOE). The left column shows INR + Spectral Embedding interpolation, providing detailed continuous spatial distribution. The middle column displays INR + Spectral Embedding + Regional Averaged results, offering smoother transitions while maintaining regional boundaries. The right column demonstrates Abagen's region-wise interpolation, which generates discrete values for brain regions. Color intensity represents expression levels, with lighter colors indicating lower expression (0.0) and darker colors indicating higher expression (1.0). \textbf{b}: Axial view comparison of gene expression interpolation for ANK3 and SIRPA, which were specifically selected due to their adjacency in the spectral embedding space, suggesting they should exhibit similar spatial expression patterns. From top to bottom: INR, INR + Spectral Embedding, INR + Spectral Embedding + Regional Averaged, and Abagen. When using spectral embedding (middle rows), we expect these genes to show highly similar spatial distributions, particularly in posterior brain regions, which is not necessarily apparent when using standard INR without embedding (top row). This similarity becomes increasingly evident moving from INR + Spectral Embedding to Regional Averaged and Abagen methods. Expression levels are color-coded from -0.2 (light) to 1.0 (dark), with brain regions delineated by gray contours. This comparison demonstrates how spectral embedding effectively captures functional relationships between genes by aligning their spatial expression patterns.}
    \label{fig:glassbrain}
\end{figure*}

\section{Results}
In this section, we describe the full set of experimental evaluations of the model. The main objective of these experiments is to determine whether the model can accurately interpolate spatial gene expression in the human brain, using sparse microarray data, and whether adding a gene-level spectral embedding enhances biological plausibility and generalization. In particular, we aim to address following questions: Using a spectral embedding of gene correlation structure, can we uncover a biologically significant order among AD risk genes? Does the INR model reliably learn and generalize spatial expression patterns across the brain? What are the differences between our methodology and the conventional \texttt{abagen} interpolation method? Can interpolated gene expression patterns from INR capture established disease-relevant spatial relationships, such as correlations with AD-related tau pathology?

\subsection{Spectral embedding of Gene correlation matrix gives an ordinal axis over genes}

We assembled the top 100 AD risk genes' microarray expression, sampled at all available sites in the brain, and computed their pairwise correlations, between all gene pairs. Heatmaps of the resulting correlation matrices are shown in Figure \ref{fig:gene_heatmaps}a,c in Supplementary Information section, for two brain donors who had microarray sampled from both hemispheres. Since the risk genes are originally obtained from the ADSP database in no specific order, the heatmaps accordingly do not show any particular pattern. In other words, we should not expect similar genes to be placed close to each other in the gene list. 
We performed spectral embedding of the gene correlation matrix (see Methods section 3) and obtained, using the first non-zero eigenvector of the matrix's Laplacian, a one-dimensional embedding of each gene in this spectral space. We sorted the genes in ascending order of this embedding. We refer to the resulting index of the sorted genes as the ordinal score $g$ in Methods. We then reordered the existing matrix to reflect the new gene ordering based on their spectral embedding values (Figure \ref{fig:gene_heatmaps}b,d). This reordering reveals a clearer structure in the heatmaps, with genes adjacent in the ordinal score axis exhibiting stronger spatial correlations with each other, as seen in the diagonally dominant pattern of the reordered matrices. We can see that the spectral embedding of gene profiles of the top 100 AD risk genes leads to a natural disease axis --  at one end, we see genes that have a strong positive spatial correlation with AD patients' tau PET scans, while at the other end are genes with a strong negative correlation with tau. 

\subsection{INR Model Characterization, fine-tuning and evaluation}
The INR model was extensively evaluated with respect to various choices regarding network size and hyper-parameters. 
We used a rigorous validation strategy, setting aside 10\% of the data for validation purposes. The dataset was partitioned into training, validation, and test sets with a ratio of 18:1:1 respectively, ensuring sufficient data for training while maintaining adequate independent sets for validation and testing. During the training process, we monitored both training and validation losses to detect potential overfitting. The model showed desirable convergence profile, with both training and validation losses consistently decreasing and ultimately converging to below $1 \times 10^{-8}$ (Figure \ref{fig:ablation}a,b) suggesting that 
our model was able to fit the spatial patterns  of gene expression data at the measured sites/locations. 
Our ablation studies (Figure \ref{fig:ablation}c,d) show that the model maintains stable performance even under increasing noise levels, with only a modest degradation in performance as noise increases, indicating that the approach is robust.

\subsection{INR gene interpolation results of selected genes}

Visualization of gene expression patterns across five genes that are highly associated with AD using three different representation methods is presented in Figure \ref{fig:glassbrain}a. The first column displays our INR model's high-resolution voxel-wise continuous interpolation results, showcasing detailed spatial variation in gene expression throughout the brain. These voxel-level visualizations reveal fine-grained expression patterns that might be missed in regional analyses, with darker colors indicating higher expression levels and lighter colors representing lower expression. The middle next presents the regional averages computed from our INR voxel-wise results, where expression values have been aggregated within anatomically defined brain regions. This regional view provides a simplified but anatomically interpretable representation of gene expression patterns, facilitating comparison with traditional analysis approaches. The last column shows \texttt{abagen}'s regional expression results, which serve as our validation baseline. The high similarity between the second and third rows (\texttt{abagen} and INR regional) across all five genes demonstrates the reliability of our interpolation method, while the last row (INR voxel) provides additional spatial detail not available in regional-level analyzes.

\subsection{Illustration of the regularizing role of gene ordinal axis}

Our goal in incorporating spectral embedding of genes into an ordinal gene score axis is to leverage it as a powerful means to constrain the INR model and to enforce regularization. To illustrate this behavior, we chose two genes that are adjacent in the spectral embedding space - \textit{ANK3} and \textit{SIRPA}. Recall, our expectation is that these two genes should be spatially correlated. When INR is trained on all genes individually, without the use of the shared gene axis given by spectral embedding between genes, the INR interpolation results in discontinuous patterns and poor alignment between genes, due to the sparse sampling in the brain. This is shown in Figure \ref{fig:glassbrain}b (top row), this time for an axial slice. When trained jointly with spectral embedding (Figure \ref{fig:glassbrain}b second row), similar expression patterns between the two genes emerge, particularly visible in the posterior brain regions. The similarity and pattern become more prominent when we take the average of voxel values of each region. The Figure \ref{fig:glassbrain}b's bottom panel shows the \texttt{abagen} region-wise interpolation result, which appears similar to the INR result averaged within each region. This observed consistency between INR and \texttt{abagen} gives us an indication that the use of the additional ordinal score within INR, using spectral embedding, is an effective way to interpolate expression across the brain, while at the same time ensuring a close alignment between functionally related genes. This augmentation strategy allows the INR model to train much more effectively on the available gene samples and thereby overcome the spatially sparse samples available in the ABA dataset. 

\begin{table}[t!]
\centering
\begin{tabular}{ccc}
\hline
\textbf{Method} & \textbf{R} & \textbf{MSE} \\
\hline
INR & 0.9430 ± 0.0268 & 0.0070 ± 0.0029 \\
INR + Spectral Embedding & 0.9016 ± 0.0145 & 0.0095 ± 0.0008 \\
INR + Spectral Embedding + Regional Averaged & \textbf{0.9833 ± 0.0047} & \textbf{0.0015 ± 0.0004} \\
\hline
\end{tabular}
\caption{Averaged Pearson Correlation (R) and Mean Squared Error (MSE) values for INR methods, computed on a voxel-wise basis across the brain to compare interpolated gene expression with \texttt{abagen}.}
\label{tab:avg_r_mse}
\end{table}

\begin{figure}[!t]
    \centering
    \includegraphics[width=\textwidth]{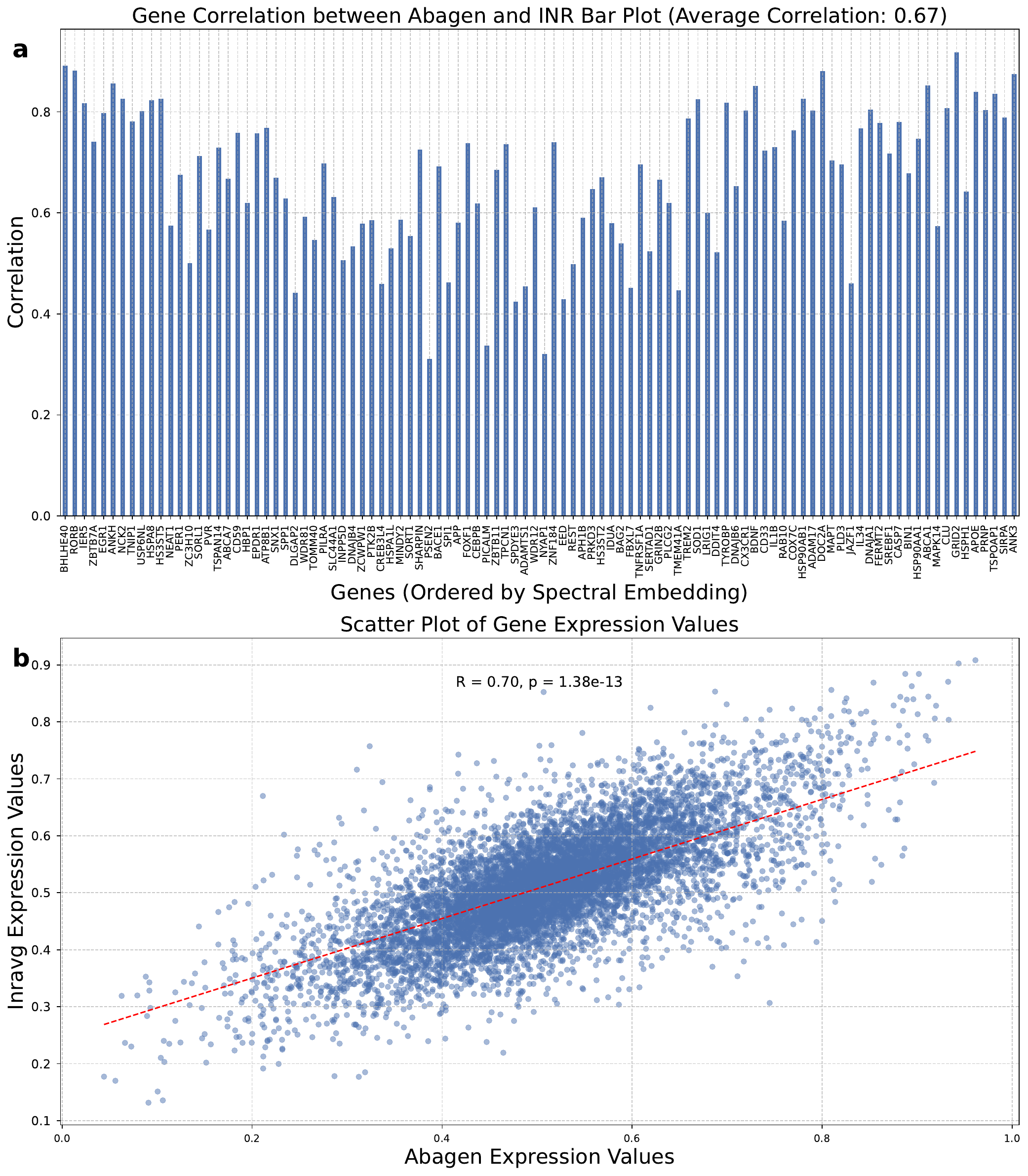}
    \caption{Region-wise comparison between INR and \texttt{abagen}. \textbf{a:} Individual genes correlation values between INR and \texttt{abagen} results ordered by spectral embedding, with an average Pearson correlation of 0.67 across all genes. \textbf{b:} Scatter plot comparing individual gene expression values between INR (y-axis) and \texttt{abagen} (x-axis) methods across all genes and all regions. Blue points represent individual expression values, with the red line showing linear regression fit ($R=0.70$, $p=1.38\times 10^{-13}$).}
    \label{fig:inr_vs_abg}
\end{figure} \noindent

\subsection{Numerical assessment of INR against \texttt{abagen}}

While cross-validation with the validation and test set are good indicators of the INR model's fit to samples, we also require external validation against a benchmark method. For this purpose, we again chose the widely-adopted \texttt{abagen} method as a comparison baseline since it provides regional interpolation based on gene probes, allowing us to assess if our interpolated gene values are reasonable.
To evaluate this, we first computed regional averages from our INR voxel-wise interpolated results and then calculated the correlation between INR and \texttt{abagen} values for each gene of interest. This analysis is a \textit{region-wise comparison}, where voxel-level predictions from INR are averaged within anatomical regions prior to comparison.
As noted above, genes were sorted using spectral embedding, and their correlations are displayed in a bar plot (Figure \ref{fig:inr_vs_abg}a). Individual genes showed varying correlation strengths, with many exhibiting strong positive correlations between 0.7 and 0.9. While this variation suggests that agreement between the two methods is gene-dependent, even the lower-performing genes maintained correlations above 0.4, indicating consistent baseline agreement. The average correlation of 0.67 across all genes (averaged over per-gene $R$'s) demonstrates robust overall agreement between INR and \texttt{abagen} approaches.
It is also worth noting that in our \textit{voxel-wise} comparison between INR and \texttt{abagen}, the INR models achieved exceptional performance, with Pearson correlation values exceeding 0.96 across genes and mean squared errors (MSE) below 0.0025. These values were computed using a \textit{voxel-wise} approach, where gene expression predictions from INR and \texttt{abagen} were directly compared at each voxel location across the brain.
This further indicates strong linear agreement in regional expression patterns and minimal absolute differences in interpolated expression levels. Averaged \textit{voxel-wise} correlation and MSE scores are provided in Table~\ref{tab:avg_r_mse}, while gene-wise \textit{voxel-wise} scores are available in Table~\ref{tab:r_mse} in the Supplementary Information section.

The scatter plot in Figure \ref{fig:inr_vs_abg}b provides a detailed visualization of the relationship between expression values from both methods, where each gene and each region is depicted by a separate circle. The plot reveals a strong positive correlation ($R=0.70$) with high statistical significance ($p=1.38\times 10^{-13}$), indicating that the agreement between methods is not attributable to chance. The data points form a dense, elongated cloud following a clear linear trend, illustrated by the red regression line. The $x$- and $y$-axes represent regional gene expression values from INR and \texttt{abagen}, respectively.
Point density is highest in the middle range of expression values, suggesting strong consistency between methods in capturing moderate expression levels. The scatter plot exhibits some heteroscedasticity, with greater variance in the middle range compared to the extremes, indicating that both methods may agree more strongly when identifying genes with very low or very high expression levels. The strong positive correlation and high statistical significance between INR and \texttt{abagen} results 
suggest that the model can accurately interpolate gene expression values across the brain.

\subsection{Assessing the relationship between regional gene expression and AD-associated regional tau uptake}

\begin{figure}[H]
    \centering
    \includegraphics[width=\textwidth]{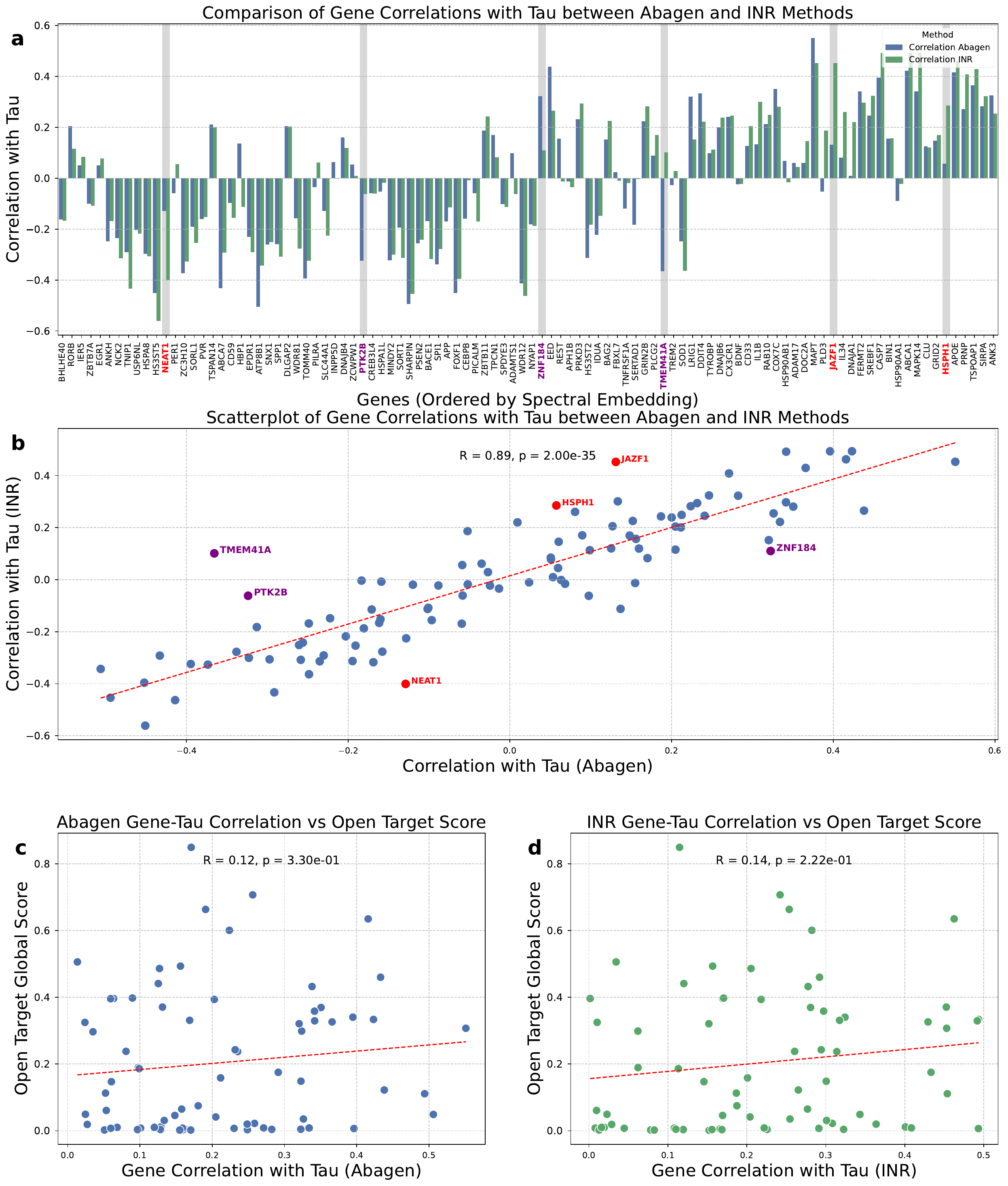}
    \caption{Comparative tau correlation analysis of gene between \texttt{abagen} and INR methodologies. \textbf{a.} Bar correlation cooefficients plot between gene expression and tau, \texttt{abagen} in blue, INR in green. Genes are ordered based on spectral embedding. \textbf{b.} Scatter plot between \texttt{abagen} and INR's correlation cooefficients, strong positive correlation ($R=0.89$, $p=2.00 \times 10^{-35}$) indicates high consistency between two methods. We highlighted three geners that had a higher association with tau under INR (\textit{NEAT1, JAZF1, HSPH1}, denoted in red) and three that had a higher association with tau under \texttt{abagen} (\textit{TMEM41A, PTK2B, ZNF184}, denoted in purple). \textbf{c.} Scatterplot comparing \texttt{abagen}-tau correlations (x-axis) with Open Target Global Scores (y-axis) ($R=0.12$, $p=0.33$). \textbf{d.} Scatterplot comparing INR-tau correlations (x-axis) with Open Target Global Scores (y-axis) ($R=0.14$, $p=0.222$).}
    \label{fig:tau_opentarget}
\end{figure}

To assess our method's relevance in the context of AD pathology, we examined the relationship between gene expression patterns and tau-PET data in Alzheimer's disease patients from the ADNI database (\url{adni.loni.usc.edu}). As noted in Methods section, this gives a regional tau vector $\mathbf{\tau} \in \mathcal{R}^N$ of the average tau uptake across the ADNI patient cohort. To ensure anatomical correctness, all tau correlation analyses were performed using region-wise gene expression values, obtained by averaging voxel-level estimates within anatomical regions. We calculated correlations between $\mathbf{\tau}$ and gene expression values derived from both INR and \texttt{abagen} methods. Figure \ref{fig:tau_opentarget} displays these correlations across AD-implicated genes, ordered by spectral embedding. The bar plot reveals varying correlation patterns, with some genes showing strong positive correlations (up to 0.5) and others exhibiting negative correlations (down to -0.5), suggesting diverse relationships between gene expression and tau pathology. Comparing correlations from both methods (INR in green, \texttt{abagen} in blue), we found that the two methods produce generally similar correlations with tau for most genes. The minor differences we noted between the two methods reflect the distinct approaches each method uses to handle spatial interpolation of gene expression data. Nonetheless, the general high agreement in correlation strength with tau from INR and \texttt{abagen} provides additional validation of the model's ability to accurately predict gene expression patterns across the brain.

The spectral embedding ordering reveals a clear gradient structure in the gene-tau relationships, with genes systematically arranged from negative to positive correlations. This organization through spectral embedding provides valuable insights into the functional relationships among genes and their potential roles in tau pathology. Genes clustered on the right side of the plot predominantly show positive correlations with tau-PET signals, suggesting these genes may be associated with increased tau accumulation or may be upregulated in response to tau pathology. Conversely, genes clustered on the left side exhibit negative correlations, indicating they might play protective roles against tau accumulation or are downregulated in regions with high tau burden. This clear separation and gradual transition in correlation values, revealed through spectral embedding, suggests the existence of distinct functional gene modules that may work in concert to either promote or protect against tau pathology. 
We also show the robust agreement quantitatively between methods in the scatter plot (Figure \ref{fig:tau_opentarget}b), which shows a strong positive correlation ($R=0.89$, $p=2.00 \times 10^{-35}$) between tau correlations derived from both methods. The low $p$-value suggests that this agreement is due to similarities in results between two methods, while the high correlation coefficient indicates strong linear relationship between the two methods' results. The scatter plot reveals several interesting features: (1) a clear linear trend that spans the entire range of correlation values, as highlighted by the red regression line, (2) relatively uniform scatter around the regression line across different correlation values, suggesting consistent agreement regardless of correlation strength, and (3) particularly tight clustering of points in the central region of the plot, indicating strong agreement for genes with moderate tau correlations.
It is also worth checking the patterns at the extremes of the distribution in the scatter plot. Both methods consistently identify the same genes as having the strongest positive and negative correlations with tau pathology, as evidenced by the points in the upper right and lower left quadrants. Some of the most important known risk genes in AD (e.g. \textit{APOE, ABCA7, ABCA1, TOMM40, SHARPIN, PSEN2, SOD1, FOXF1, MAPT, MAPK14, PRNP, TSPOAP1}) give generally high correlations with the regional distribution of AD tau under both methods of interpolation. 
Despite the general agreement between the two methods, there were some important discrepancies. We identified six genes whose correlation with tau was especially discordant between INR and \texttt{abagen} (Figure \ref{fig:tau_opentarget}a,b): NEAT1, JAZF1, HSPH1 (higher association with tau under INR, denoted in red) and TMEM41A, PTK2B, ZNF184 (higher association with tau under \texttt{abagen}, denoted in purple). For this purpose we disregarded the sign of the correlation. 

\FloatBarrier
 \noindent

\subsubsection{Open Target Database Correlation Analysis}

Although our finding that gene-tau correlations are strong for some genes and show a gradient along the spectral embedding axis, it is important not to over-interpret these correlations as evidence for a given gene's direct role in AD pathophysiology. While some correlation with the spatial patterning of tau and genes should be expected in this set of AD risk genes, our analysis uses the genes' healthy expression in the brain, rather than disease-associated dysregulation. 
This impression is confirmed by our final analysis, where we compared the gene-tau correlation results against Alzheimer's disease-specific genetic association scores from the Open Targets database \cite{koscielny2017open}, which integrates genetic and disease data from multiple sources. Figure \ref{fig:tau_opentarget}c and \ref{fig:tau_opentarget}d show that both interpolation methods exhibited weak positive correlations with Open Targets AD-specific scores ($R=0.12$, $p=0.330$ for \texttt{abagen}; $R=0.14$, $p=0.222$ for INR). This suggests that across all genes, spatial correlation to tau pathology does not necessarily imply direct genetic risk for AD.

\section{Discussion}

We proposed here a novel super-resolution technique to infer voxel-wise gene expression data in the human CNS from a limited number of microarray samples. The method is based on the deep neural network architecture called implicit neural representations (INR), but with domain-specific customizations and augmentation necessary for the current task. Since there is no ground truth available for ABA gene expression we instead rely on indirect assessments to demonstrate the concept validity of the proposed technique. In our current setup, we chose 100 genes that are known to be involved with Alzheimer's disease, and perform brain-wide interpolation of these genes. Our reasoning is threefold: First, we wish to produce gene maps that hold some clinical or scientific value, in this case, due to their association with a known and important neurological disease. Second, we wish to constrain the mapped genes to vary along a familiar gene similarity axis. Since there are ~22000 genes, and most of these do not have brain-specific expression, it is important to constrain the list of genes to a brain-specific process - in this case their association with AD. Third and most important, the gene axis, obtained by spectral embedding, served as an augmented ordinal index that was used by the INR model to enrich its training features and to unlock more powerful non-local learning. We note that the technique is completely agnostic to the specific choice of genes and can be easily extended to other disease contexts. Our approach can therefore be broadly useful for researchers interested in exploring ST in the human brain. INR represents a significant advance in producing super-resolved spatial gene expression, with broad applicability in spatial transcriptomics, neurodegenerative disease and other fields of neuroscience. Our main results and their implications are discussed below. 

\subsection{Major findings}

First, we showed that spectral embedding of gene profiles of the top 100 AD risk genes leads to a natural disease axis, whereby at one end are genes that have a strong positive spatial correlation with AD patients' tau PET scans, while at the other end are gens with a strong negative correlation with tau (Figure \ref{fig:tau_opentarget}a). Each gene on this axis resembles its neighboring genes in spatial distribution; hence the gene axis serves as a powerful means to constrain the INR model and to enforce regularization. 

Second, we showed that this augmentation strategy allows the INR model to train much more effectively on the available gene samples and thereby overcome the spatially sparse samples available in the ABA dataset. Without this augmentation, INR was less successful in producing plausible spatial distributions of genes, while its inclusion led to gene maps that were both spatially smooth as well as consistent between neighboring genes (Figure \ref{fig:glassbrain}). 

Third, after averaging our voxel-wise gene maps into regional values based on the 86-region Desikan-Killiany atlas parcellation, we could directly compare our outputs with the region-level maps obtained previously by a well-established and widely used interpolation tool called \texttt{abagen} \cite{Markello_abagen_2021}. We showed that INR-estimated expression matched \texttt{abagen} interpolation at the regional level with a strong and highly significant correlation (Figure \ref{fig:inr_vs_abg}). We report that the average correlation between the two methods was $R=0.67$ (average of R per gene) and $R=0.70$ (aggregate R across all genes and regions). 

Fourth, we demonstrated specific neurological context in which the proposed method can be applied. For this purpose we chose a large public Alzheimer's disease dataset of patients' tau-PET scans. On this data we showed that INR maps possess very strong and clear spatial association with AD pathologic hallmark, specifically, tau-PET uptake. Nonetheless, INR-inferred AD risk gene expression showed highly variable associations with tau-PET data in ADNI patients, both between patients and genes (Figure \ref{fig:tau_opentarget}a,b). Yet again, the correlations we report between INR and AD tau are similar to those between \texttt{abagen} and tau, with $R=0.89$ between the two methods, confirming that the INR method produces results with consistency to prior approaches at the regional level. Some of the most important known risk genes in AD (e.g. \textit{APOE, ABCA7, ABCA1, TOMM40, SHARPIN, PSEN2, SOD1, FOXF1, MAPT, MAPK14, PRNP, TSPOAP1}) give generally high correlations with the regional distribution of AD tau under both methods of interpolation. INR-based approach reliably captures pathologically significant relationships while providing enhanced spatial resolution through continuous interpolation. This enhanced resolution could be particularly valuable for studying specific brain regions where tau pathology is known to initiate or progress rapidly during AD progression. This is particularly important for identifying potential therapeutic targets, as genes with strong correlations might play crucial roles in tau-related pathological processes.

Fifth and notwithstanding the previous point, we found certain intriguing points of departure between INR and \texttt{abagen} outcomes that we believe are relevant in the AD context. We highlighted six genes whose correlation with tau was especially discordant between INR and \texttt{abagen} (see Figure \ref{fig:tau_opentarget}a,b): \textit{TMEM41A, JAZF1, HSPH1, NEAT1, PTK2B, ZNF184}. A summary of their biological pathways and AD associations is given below, that together serve to highlight that these points of departure between INR and \texttt{abagen} can lead to quite different biological conclusions and novel insights previously inaccessible to the community. 

Together, these data support our goal of delivering high-quality and fully quantitative gene expression maps with a spatial resolution that exceeds the limit imposed by the sparsely sampled raw microarray data. 

\subsection{Biological pathways and disease association of selected discordant genes}

Although we do not have access to ground truth gene expression in the human CNS, we surmised that the ``better'' method would produce more biologically plausible gene associations with AD tau. Among the genes that were the most discordant between INR and \texttt{abagen}, here we selected three that had a higher association with tau (irrespective of sign) under INR (\textit{NEAT1, JAZF1, HSPH1}, denoted in red in Figure \ref{fig:tau_opentarget}a,b) and three that had a higher association with tau (irrespective of sign) under \texttt{abagen} (\textit{TMEM41A, PTK2B, ZNF184}, denoted in purple). We did not include in this section those genes that had a modest association with tau under both INR and \texttt{abagen}, even if they showed a large discrepancy between the two methods. We relied on the public Open Targets database \cite{koscielny2017open} to obtain information about pathways and AD association of these selected genes, as summarized below.

\textbf{\textit{NEAT1}}: Nuclear Paraspeckle Assembly Transcript 1 is not highly expressed in the brain, nor appears to have a strong association with AD pathophysiology. It has been associated instead with Dengue Disease and Gastric Adenocarcinoma. That INR produced a strong correlation ($R=-0.40$) while \texttt{abagen} gave an insignificant one, suggests that in this case the latter may be closer to biological plausibility.

\textbf{\textit{JAZF1}}: \textit{JAZF1} is implicated in type-2 diabetes (T2D) in addition to AD, and our analysis showed that its regional expression levels were highly correlated with residual levels of tau, suggesting its plausible function as an AD risk factor. Due to the link between T2D and AD \cite{2004_Janson_T2D}, \textit{JAZF1} poses an excellent candidate gene open to further exploration as an AD risk gene. The fact that INR gives one of the highest correlations between this gene and tau, while \texttt{abagen} gives only a moderate correlation, is an indication that perhaps INR is capturing its role in AD pathophysiology better than \texttt{abagen} is. 

\textbf{\textit{HSPH1}}: This heat shock protein family member gene acts as a nucleotide-exchange factor (NEF) for chaperone proteins HSPA1A and HSPA1B, promoting the release of ADP from HSPA1A/B thereby triggering client/substrate protein release. Prevents the aggregation of denatured proteins in cells under severe stress, on which the ATP levels decrease markedly. It is highly expressed in the brain, and its family member genes are highly implicated in amyloidosis, a key process in AD pathophysiology. Although INR produced only a moderate correlation with tau ($R=0.29$), this was much higher than that produced by \texttt{abagen} ($R=0.05$). 

\textbf{\textit{TMEM41A}}: The Transmembrane protein 41A gene is not particularly expressed in the CNS, and does not appear to have a strong link to AD, as per Open Targets platform. Yet this gene was given a highly significant negative correlation with tau ($R=-0.36$) by \texttt{abagen}, while INR only produced an insignificant correlation with tau ($R=0.1$). In this case, we may conclude that INR values may be closer to biological plausibility than \texttt{abagen}'s.

\textbf{\textit{PTK2B}}: Protein Tyrosine Kinase 2 Beta has been reported to become aberrant in cancer cell invasion, in tumor formation and metastasis. Elevated PTK2B/PYK2 expression is seen in gliomas, hepatocellular carcinoma, lung cancer and breast cancer. However it does not seem to have a strong association with AD. Here \texttt{abagen} produced a strong negative correlation ($R=-0.32$) while INR gave an insignificant one, suggests that in this case INR may be closer to biological plausibility than \texttt{abagen}.

\textbf{\textit{ZNF184}}: Zinc Finger Protein 184 is a Protein Coding gene that is expressed in the brain. Diseases associated with ZNF184 include Prostate Cancer and Renal Pelvis Transitional Cell Carcinoma. No strong association with AD was reported. That \texttt{abagen} produced a strong correlation with AD tau ($R=0.32$) while INR gave a weaker one ($R=0.10$) supports the latter as being more biologically plausible.

We conclude from this analysis that the genes whose correlation with tau was especially inconsistent between INR and \texttt{abagen} had biological and disease pathways that broadly supported INR over \texttt{abagen}. This conclusion is only suggestive and anecdotal rather than comprehensive, yet it points to the fact that the biological or disease role assigned to a given gene in ST studies can change dramatically depending on the method used to interpolate or super-resolve them from sparse sampled microarray data. We also assessed as a final step the relationship between inferred genes' correlation with tau and their known genetic risk score tabulated by Open Targets database (Figure \ref{fig:tau_opentarget}c,d). We did not expect that the two metrics should be strongly correlated; yet we would expect at least a moderate correlation, given that the genes involvement in the disease may be reflected in their spatial association with pathology in the CNS. Our results are in line with this reasoning; however, we did not see a meaningful difference between INR and \texttt{abagen} on this account ($R=0.14$ vs $R=0.12$).

\subsection{Potential applications in neuroscience}

We anticipate that our voxel-level maps will become a critical tool for assessing the role of ST across a variety of applications. We discuss below some of these in context of current literature.

\subsubsection*{Spatial transcriptomics can reveal molecular correlates of brain structure and function}

The structural connectome, which represents the density of physical projections between brain regions and is measured by such techniques as viral tracing and diffusion tensor imaging, is a coarse wiring diagram of the central nervous system (CNS) \cite{sporns2005human, oh2014mesoscale, bullmore2011brain, zeng2018mesoscale}. 
Complex molecular processes during embryonic development encourage the formation of connections between brain regions, and later postnatal pruning results in structural connectomes with a remarkable degree of conservation between healthy individuals. Several  groups report that connected regions tend to have higher correlated gene expression patterns than regions that are not \cite{french2011large, ji2014integrative}. There is accordingly a strong interest in gaining a rigorous measure of how gene expression and cell type composition of brain regions relate to connectivity\cite{french2011large, tan2013neuron}, which can deepen our understanding of how brain circuits mature during the development of the CNS and how they are disrupted in neurodegenerative diseases, among other areas of inquiry. The correlation between regional gene expression and connectivity is well established in mice\cite{french2011large, henriksen2016simple, fulcher2016transcriptional, reimann2019null} and humans\cite{goel2014spatial, vertes2016gene, diez2018neurogenetic}. French, \textit{et al.} built statistical models correlating the gene expression signatures of 17,530 genes in 142 anatomical regions from the \noindent Allen Brain Atlas, and identified a subset of genes that are statistically correlated with the brain's wiring diagram\cite{french2011large}. A similar approach was applied in humans, and shown to give strong evidence that gene expression in CNS is predictive of long-range anatomic connectivity measured from diffusion-weighted MRI scans \cite{goel2014spatial}. Recently, a machine learning study from our laboratory established the strong ability of cell type composition, in turn derived from spatial gene expression, in predicting the short- and long-range connectivity in the mouse brain \cite{}. Further advances in these types of explorations would require ever-improving maps of gene expression in the CNS, especially in human brains where experimental measurements are necessarily far more challenging and limiting.

\subsubsection*{From genes to cells}

Characterizing whole-brain distributions of neural cell types is a topic of keen interest in modern neuroanatomy, with many applications to both basic and clinical neuroscience research. Advances in deconvolution and clustering algorithms applied to gene expression data are enabling quantitative brain-wide neuronal and nonneuronal cell–type density maps; see e.g. the MISS technique from our group \cite{mezias2022matrix}. However, unlike animal studies above, the quantification of human cell type densities will require higher-resolution gene expression than currently available; this represents a key motivation behind the current study.

\subsubsection*{Utility of spatial transcriptomics in neurodegenerative diseases}

The understanding of how local/regional genetic factors govern or mediate the selective vulnerability of certain brain areas to pathological insults is a central goal of neurodegenerative research.
Although many risk genes in Alzheimer's and other dementias are known from genome- and transcriptome-wide association studies \cite{sun2021transcriptome, seshadri2010genome, bellenguez2022new, kunkle2019genetic}, it is still not known whether and how the presence of certain molecular factors and their gene regulators at \textit{baseline} (that is, in the non-disease condition) may confer inherent vulnerability or resilience even before the establishment of overt pathology.
Prior analyses have yielded conflicting results, pointing to a well-known and puzzling spatial dissociation between upstream risk genes and downstream disease topography \cite{Fusco_1999, Subramaniam_2019}. Instead, the concerted action of large number of risk genes would be a far better neural correlate of these disorders. Recent work has begun to elucidate the complex relationships between gene transcription at a whole-brain level and AD-associated pathology \cite{sepulcre_neurogenetic_2018, anand2024selective}, but further experimental and clinical work is required to validate these associational studies. This represents a crucial opportunity for ST in understanding regional vulnerability. By producing a super-resolved map of gene expression from the established AHBA, we plan to complement this work at a whole-brain level at a spatial resolution previously inaccessible. A full exposition of candidate genes and their associated functional networks may help the field of AD in its quest to better understand mechanisms of disease and find new drug targets. 

\subsection{Limitations and future directions}
An important limitation of our study is that gene expression data from AHBA are from only six unique healthy subjects, only two of whom had bilaterally sampled gene expression~\cite{Hawrylycz_atlas_human_brain_transcriptome_2012}.
While this impacts the generalizability of the gene maps produced by our approach, it is a limitation that is common to brain-wide transcriptomic studies in humans to date, and no alternate datasets exist in the public domain with greater spatial resolution or sampling density.
As and when these data become public, our approach will be able to incorporate them and produce ever-improving maps, without significantly changing the underlying model.
Further, while INR is a robust technique that has enjoyed success for a variety of problems \cite{chen2021learning, sun2023next3d, yu2021pixelnerf, niemeyer2020differentiable, mildenhall2020nerf}, voxel-level mapping of gene expression is difficult to validate, as there is no ground truth against which to compare our results.
Supporting the validity of our method is that group averaging of region voxels accurately reproduces \texttt{abagen}-inferred values on the DK atlas.

We note that alternative designs could potentially lead to faster convergence, improved generalization, and a reduction in the number of parameters that need to be optimized. A promising formulation is the Operator-Implicit Neural Representation (O-INR) \cite{pal2024implicit}. Unlike traditional models that map positional encodings to a signal, O-INR operates by mapping one function space to another. For gene expression data, these function spaces can be well described along the previously introduced spectral embedding dimension, and can enable more efficient and compact training, particularly for learning complex signals. Future work will extend the current model to incorporate O-INR components to evaluate potential improvements.


Future work includes using INR on other datasets that require interpolation to be extended to the whole brain.
Recently, the Brain Initiative Cell Census Network produced a transcriptomic atlas of cell types in the healthy human brain, sampling from over 100 discrete sites and profiling over 3000 unique kinds of neuronal and non-neuronal cells \cite{siletti2023transcriptomic}. 
This dataset should provide sufficient spatial resolution for us to apply INR directly, resulting in a spatial atlas of these cell types at a voxel-wise level.
Similarly, we expect that we can use INR on data derived from PET imaging, such as those explored here for tau, which lack the spatial resolution of MRI and exhibit a low signal-to-noise ratio in individual voxels.
This will facilitate a more detailed investigation of Alzheimer's disease using datasets that have already been produced through consortia such as ADNI, among other neurodegenerative diseases.

%
%
%
\section{Data availability} 
We plan to share all relevant data (regional gene expression, tau) and source code publicly upon paper acceptance, via our GitHub repository (\url{https://github.com/vsingh-group/gene-expression-inr}). Original gene expression data may be obtained directly from the ABI.
\section*{Acknowledgements}
We thank the ABI for making their gene data available publicly and to ADNI for the AD-related tau data. 

\section*{Funding}
This study was partially supported by NIH grants R01NS092802, RF1AG062196, and R01AG072753 awarded to Ashish Raj.

\section*{Competing interests}
The authors have no competing financial or other interests.

\clearpage

\section*{Author contributions statement}
V.S. and A.R. conceived the study and jointly supervised its execution. X.Y., S.P. and J.T. conducted the literature search, analyzed and processed the data. X.Y. implemented the INR model and produced result figures and tables. All authors contributed to drafting and refining the manuscript. J.T. and S.P. helped with data analysis. All authors reviewed the and approved the final manuscript.



\beginsupplement

\begin{samepage}

\section*{Supplementary Information}

This section contains supplementary methods, algorithm pseudocode, supplementary figures and tables. 

\subsection*{Supplementary Methods}  

\subsubsection*{Pseudocode of INR algorithm}



\begin{algorithm}[H]
    \caption{INR Interpolation Pipeline for Brain Gene Expression Analysis}
    \begin{algorithmic}[1]
    \Require $\mathbf{A}$: Allen Human Brain Atlas data
    \Require $\mathbf{G} = \{g_1, g_2, ..., g_k\}$: Set of AD-related gene names
    \Function{GenerateEmbeddings}{$\mathbf{G}$}
        \State $c \gets$ ClassifyMatter($\mathbf{G}$) \Comment{$c \in \{-1, 0, 1\}$}
        \State $\phi \gets$ ComputeSpectralEmbedding($\mathbf{G}$) \Comment{$\phi \in \mathbb{R}^d$} 
        \State $\mathbf{f} \gets$ $[c; \phi]$ \Comment{Feature concatenation}
        \State \Return $\mathbf{f}$
    \EndFunction
    \Function{PreprocessData}{$\mathbf{A}, \mathbf{G}$}
        \State $\mathbf{X} \in \mathbb{R}^{k \times n}$ $\gets$ FilterGenes($\mathbf{A}, \mathbf{G}$) \Comment{n $\approx$ 62000 probes}
        \State $\mathbf{X} \gets$ NormalizeData($\mathbf{G}, \mathbf{X}$)
        \State $\mathbf{X} \gets$ GenerateEmbeddings($\mathbf{X}$)
        \State \Return $\mathbf{X}$
    \EndFunction
    
    \Function{Training}{$\mathbf{X}$}
        \State $\theta \gets$ InitializeParameters() \Comment{Neural network parameters}
        \State $L \gets 0$ \Comment{Loss initialization}
        \For{$t = 1$ to $T$} \Comment{Training iterations}
            \ForAll{$\mathbf{x} \in \mathbf{X}$} \Comment{$\mathbf{R}$: set of brain regions}
                \State $\mathbf{e} \gets$ PositionalEmbedding($\mathbf{x}$)
                \State $(x, y, z, c, \phi, \mathbf{e}) \gets \mathbf{x}$ \Comment{$\mathbf{e}$: embedding features}
                \State $v_{true} \gets$ GetActualGeneExpression($\mathbf{x}$)
                \State $v_{pred} \gets \text{INR}_\theta(\mathbf{x})$ \Comment{Forward pass}
                \State $L \gets L + \|v_{pred} - v_{true}\|_2^2$
            \EndFor
            \State $\theta \gets \theta - \eta\nabla_\theta L$ \Comment{Parameter update}
        \EndFor
        \State \Return $\text{INR}_\theta$
    \EndFunction

    \Function{Inference}{$\text{INR}_\theta, c, \phi$}
        \State $X, Y, Z \gets$ NiftiShape() \Comment{Get 3D volume dimensions}
        \State $V \gets \text{zeros}(X, Y, Z)$ \Comment{Initialize output volume}
        \For{$x = 1$ to $X$}
            \For{$y = 1$ to $Y$}
                \For{$z = 1$ to $Z$}
                    \State $\mathbf{e} \gets$ PositionalEmbedding($x, y, z$)
                    \State $\mathbf{x} \gets [x, y, z, c, \phi, \mathbf{e}]$ \Comment{Construct input feature}
                    \State $v \gets \text{INR}_\theta(\mathbf{x})$ \Comment{Predict gene expression}
                    \State $V[x,y,z] \gets v$ \Comment{Store prediction}
                \EndFor
            \EndFor
        \EndFor
        \State \Return $V$ \Comment{Return 3D volume of gene expression values}
    \EndFunction
    
    \Procedure{Main}{}
        \State $\mathbf{X} \gets$ PreprocessData($\mathbf{A}, \mathbf{G}$)
        \State $\text{INR}_\theta \gets$ Training($\mathbf{X}$)
        \For{$g$ in $\mathbf{G}$}
            \State $V_g \gets$ Inference($\text{INR}_\theta, \phi_g, c$)
        \EndFor
    \EndProcedure
    \end{algorithmic}
\label{algo:flow}
\end{algorithm}

\end{samepage}

\clearpage

\subsection*{Supplementary Figures and Tables}

\begin{table}[htbp]
\small
\centering
\setlength{\tabcolsep}{3pt}
\begin{tabular}{@{\hspace{2pt}}*{8}{c}@{}}
\toprule
\multicolumn{2}{c}{1-25} & \multicolumn{2}{c}{26-50} & \multicolumn{2}{c}{51-75} & \multicolumn{2}{c}{76-100} \\
\cmidrule(r){1-2}\cmidrule(r){3-4}\cmidrule(r){5-6}\cmidrule(r){7-8}
Gene & Spectral Embedding & Gene & Spectral Embedding & Gene & Spectral Embedding & Gene & Spectral Embedding\\
\midrule
BHLHE40 & -0.0462 & WDR81 & -0.0261 & NYAP1 & -0.0113 & RAB10 & 0.0291 \\
RORB & -0.0447 & TOMM40 & -0.0246 & ZNF184 & -0.0095 & COX7C & 0.0301 \\
IER5 & -0.0444 & PILRA & -0.0245 & EED & -0.0082 & HSP90AB1 & 0.0314 \\
ZBTB7A & -0.0439 & SLC44A1 & -0.0232 & REST & -0.0036 & ADAM17 & 0.0337 \\
EGR1 & -0.0436 & INPP5D & -0.0229 & APH1B & -0.0034 & DOC2A & 0.0346 \\
ANKH & -0.043 & DNAJB4 & -0.0225 & PRKD3 & -0.0033 & MAPT & 0.0348 \\
NCK2 & -0.0419 & ZCWPW1 & -0.0224 & HS3ST2 & -0.0015 & PLD3 & 0.0364 \\
TNIP1 & -0.0378 & PTK2B & -0.022 & IDUA & -0.0008 & JAZF1 & 0.0399 \\
USP6NL & -0.0378 & CREB3L4 & -0.0218 & BAG2 & 0.0 & IL34 & 0.0405 \\
HSPA8 & -0.037 & HSPA1L & -0.0208 & FBXL7 & 0.0016 & DNAJA1 & 0.0431 \\
HS3ST5 & -0.0355 & MINDY2 & -0.0207 & TNFRSF1A & 0.0068 & FERMT2 & 0.0436 \\
NEAT1 & -0.0354 & SORT1 & -0.0206 & SERTAD1 & 0.0076 & SREBF1 & 0.0465 \\
PER1 & -0.0351 & SHARPIN & -0.0204 & GRIN2B & 0.008 & CASP7 & 0.0471 \\
ZC3H10 & -0.0349 & PSEN2 & -0.0195 & PLCG2 & 0.01 & BIN1 & 0.0494 \\
SORL1 & -0.034 & BACE1 & -0.0194 & TMEM41A & 0.0103 & HSP90AA1 & 0.0496 \\
PVR & -0.0327 & SPI1 & -0.0185 & TREM2 & 0.0141 & ABCA1 & 0.051 \\
TSPAN14 & -0.0327 & APP & -0.0185 & SOD1 & 0.0144 & MAPK14 & 0.0596 \\
ABCA7 & -0.0309 & FOXF1 & -0.0181 & LRIG1 & 0.0155 & CLU & 0.064 \\
CD59 & -0.0305 & CEBPB & -0.017 & DDIT4 & 0.0202 & GRID2 & 0.0662 \\
HBP1 & -0.0304 & PICALM & -0.0169 & TYROBP & 0.024 & HSPH1 & 0.0668 \\
EPDR1 & -0.0296 & ZBTB11 & -0.0168 & DNAJB6 & 0.0256 & APOE & 0.0692 \\
ATP8B1 & -0.0295 & TPCN1 & -0.0166 & CX3CR1 & 0.0257 & PRNP & 0.0721 \\
SNX1 & -0.0287 & SPDYE3 & -0.0165 & BDNF & 0.0263 & TSPOAP1 & 0.0734 \\
SPP1 & -0.0286 & ADAMTS1 & -0.0141 & CD33 & 0.0272 & SIRPA & 0.0745 \\
DLGAP2 & -0.027 & WDR12 & -0.0138 & IL1B & 0.0275 & ANK3 & 0.0759 \\
\bottomrule
\end{tabular}
\caption{Gene names and their corresponding Spectral Embedding values}
\label{tab:gene_names_100_se}
\end{table}

\begin{table}[H]
\centering
\begin{tabular}{ll}
\hline
{\textbf{Gene symbol}} & {\textbf{Gene name}} \\
\hline
       ABCA1   &       ATP binding cassette subfamily A member 1   \\                  
       ANK3    &       Ankyrin 3                                    \\                 
       APOE    &       Apolipoprotein E                                           \\
       BAG2    &       BCL2 associated gene cochaperone                           \\
       BDNF    &       Brain-derived neurotrophic factor                          \\
       BIN1    &       Bridging integrator 1                                      \\
       CASP7   &       Caspase 7                                                  \\
       CD33    &       Siglec-3                                                   \\
       CX3CR1  &       C-X3-C motif chemokine receptor 1                          \\
       DDIT4   &       DNA damage inducible transcript 4                          \\
       DNAJB4  &       DnaJ heat shock protein family (Hsp40) member B4           \\
       FERMT2  &       Fermitin family homolog 2                                  \\
       GRID2   &       Glutamate ionotropic receptor delta type subunit 2         \\
       GRIN2B  &       Glutamate receptor ionotropic, NMDA 2B                     \\
       HSP90AB1 &       Heat shock protein 90 alpha family class B member 1       \\
       IL1B    &       Interleukin-1beta                                          \\
       IL34    &       Interleukin-34                                             \\
       LRIG1   &       Leucine-rich repeats and immunoglobulin-like domain protein 1 \\
       MAPT    &       Microtubule associated protein tau                         \\
       PLCG2   &       Phospholipase C gamma 2                                    \\
       PRKD3   &       Serine/threonine-protein kinase D3                         \\
       SREBF1  &       Sterol regulatory element binding transcription factor 1   \\
       TNFRSF1A &       Tumor necrosis factor receptor superfamily 1A             \\
       TSPOAP1 &       TSPO associated protein 1                                  \\
       ZNF184  &       Zinc finger protein 184                                    \\
       
CLU    &        Clusterin                                       \\
COX7C  &        Cytochrome c oxidase subunit 7C                 \\
DLGAP2 &        Disks large-associated protein 2                \\
DNAJA1 &        DnaJ heat shock protein family (Hsp40) member A1 \\
DNAJB6 &        DnaJ heat shock protein family (Hsp40) member B6 \\
DOC2A  &        Double C2 domain alpha                          \\
EED    &        Embryonic ectoderm development                  \\
HSPH1  &        Heat shock protein family H (Hsp110) member 1  \\
JAZF1  &        Juxtaposed with another zinc-finger 1           \\
MAPK14 &        Mitogen-activated protein kinase 14             \\
PLD3   &        Phospholipase D3                                \\
PRNP   &        Prion protein                                   \\
RORB   &        Retinoid-related orphan nuclear receptor beta1  \\
SIRPA  &        Signal regulatory protein alpha                 \\
TSPAN14 &       Tetraspanin-14                                  \\
ZBTB11 &        Zinc finger and BTB domain containing 11       \\
\bottomrule
\end{tabular}
\caption{List of AD-related risk genes used in this study}
\label{tab:gene_names}
\end{table}

\begin{table}[H]
\ContinuedFloat
\centering
\begin{tabular}{ll}
\hline
{\textbf{Gene symbol}} & {\textbf{Gene name}} \\
\hline
       ADAM17    &     ADAM metallopeptidase domain 17                         \\
       ADAMTS1   &     ADAM metallopeptidase with thrombospondin type 1 motif 1 \\
       ANKH      &    ANKH inorganic pyrophosphate transport regulator          \\
       APP       &     Amyloid beta precursor protein                           \\
       ATP8B1    &     ATPase phospholipid transporting 8B1                     \\
       BACE1     &     Beta secretase                                           \\
       BHLHE40   &     Basic helix-loop-helix family member E40                 \\
       CEBPB     &     CCAAT enhancer-binding protein beta                     \\
       EGR1      &     Early growth response 1                                 \\
       FBXL7     &     F-box and leucine rich repeat protein 7                 \\
       FOXF1     &     Forkhead box protein F1                                 \\
       HS3ST2    &     Heparan sulfate glucosamine 3-O-sulfotransferase 2      \\
       HS3ST5    &     Heparan sulfate glucosamine 3-O-sulfotransferase 5      \\
       HSPA1L    &     Heat shock 70 kDa protein 1-like                        \\
       HSPA8     &     Heat shock protein family A (Hsp70) member 8            \\
       IER5      &     Immediate early response gene 5                         \\
       NCK2      &     Non-catalytic region of tyrosine kinase adaptor protein 2 \\
       NEAT1     &     Nuclear enriched abundant transcript 1                  \\
       PER1      &     Period circadian regulator 1                            \\
       PICALM    &     Phosphatidylinositol binding clathrin assembly protein   \\
       PILRA     &     Paired immunoglobin like type 2 receptor alpha           \\
       PTK2B     &     Protein-tyrosine kinase 2-beta                          \\
       RAB10     &     Ras associated protein                                   \\
       SHARPIN   &     SHANK associated RH domain interactor                    \\
       SLC44A1   &     Choline transporter-like protein 1                       \\
       SOD1      &     Superoxide dismutase type 1                              \\
       SORL1     &     Sortilin related receptor 1                              \\
       SORT1     &     Sortilin                                                 \\
       SPDYE3    &     Speedy/RINGO cell cycle regulator family member E3       \\
       SPI1      &     Spi-1 proto-oncogene                                     \\
       SPP1      &     Secreted phosphoprotein 1                                \\
       TMEM41A   &     Transmembrane protein 41A                                \\
       TNIP1     &     TNFAIP3 interacting protein 1                            \\
       TOMM40    &     Translocase of outer mitochondrial membrane 40           \\
       TREM2     &     Triggering receptor expressed on nyeloid cells 2         \\
       WDR12     &     WD repeat domain 12                                      \\
       WDR81     &     WD repeat domain 81                                      \\
       ZBTB7A    &     Zinc finger and BTB domain containing 7A                 \\
       ZC3H10    &     Zinc finger CCCH domain-containing protein 10            \\
       ZCWPW1    &     Zinc finger CW-type and PWWP domain containing 1         \\
\bottomrule
\end{tabular}
\caption{List of AD-related risk genes used in this study (Continued)}
\end{table}

\begin{table}[H]
\ContinuedFloat
\centering
\begin{tabular}{ll}
\hline
{\textbf{Gene symbol}} & {\textbf{Gene name}} \\
\hline
       ABCA7   &       ATP binding cassette subfamily A member 7               \\
       APH1B   &       Aph-1 homolog B, gamma-secretase subunit                 \\
       CD59    &       Codes for membrane inhibitor of reactive lysis or protection \\
       CREB3L4 &       CAMP responsive element binding protein 3 like 4'        \\
       EPDR1   &       Ependymin-related protein                                \\
       HBP1    &       HMG-box transcription factor 1                           \\
       HSP90AA1 &       Heat shock protein 90 alpha family class A member 1     \\
       IDUA    &       Alpha-L-iduronidase                                      \\
       INPP5D  &       Inositol polyphosphate-5-phosphatase D                   \\
       MINDY2  &       MINDY lysine 48 deubiquitinase 2                         \\
       NYAP1   &       Neuronal tyrosine phosphorylated phosphoinositide-3-kinase \\
       PSEN2   &       Presenilin 2                                             \\
       PVR     &       Poliovirus receptor                                      \\
       REST    &       RE1 silencing transcription factor                       \\
       SERTAD1 &       SERTA domain-containing protein 1                        \\
       SNX1    &       Sorting nexin 1                                          \\
       TPCN1   &       Two pore segment channel 1                               \\
       TYROBP  &       TYRO protein tyrosine kinase-binding protein             \\
       USP6NL  &       USP6 N-terminal like        \\                   
\bottomrule
\end{tabular}
\caption{List of AD-related risk genes used in this study (Continued)}
\end{table}

\begin{table}[htbp]
\small
\centering
\setlength{\tabcolsep}{3pt}
\begin{tabular}{@{}l@{\hspace{2pt}}*{6}{c}@{}}
\toprule
& \multicolumn{2}{c}{INR} & \multicolumn{2}{c}{INR + Spectral Embedding} & \multicolumn{2}{c}{\makecell{INR + Spectral Embedding + \\ Regional Averaged}}\\
\cmidrule(r){2-3}\cmidrule(r){4-5}\cmidrule(r){6-7}
Gene & R & MSE  & R & MSE  & R & MSE\\
\midrule
TPCN1 & 0.921723 & 0.007542 & 0.904512 & 0.009471 & 0.982911 & 0.001556 \\
CEBPB & 0.950129 & 0.005371 & 0.904259 & 0.010067 & 0.976084 & 0.002256 \\
FOXF1 & 0.916187 & 0.008746 & 0.890575 & 0.009702 & 0.980469 & 0.001500 \\
FERMT2 & 0.883268 & 0.008685 & 0.887487 & 0.010118 & 0.968694 & 0.002473 \\
TSPAN14 & 0.968242 & 0.006539 & 0.905954 & 0.009166 & 0.988656 & 0.000974 \\
IDUA & 0.957360 & 0.006277 & 0.902254 & 0.008756 & 0.985089 & 0.001207 \\
APH1B & 0.952863 & 0.005655 & 0.902216 & 0.008546 & 0.983571 & 0.001337 \\
RORB & 0.953925 & 0.011347 & 0.935341 & 0.009578 & 0.988607 & 0.002052 \\
ZC3H10 & 0.965322 & 0.004844 & 0.890666 & 0.009551 & 0.984634 & 0.001186 \\
CD33 & 0.840707 & 0.010868 & 0.864247 & 0.010621 & 0.983195 & 0.001124 \\
CD59 & 0.965093 & 0.003522 & 0.898911 & 0.010012 & 0.982646 & 0.001507 \\
TSPOAP1 & 0.904637 & 0.008262 & 0.918823 & 0.008231 & 0.988381 & 0.001057 \\
IL34 & 0.906973 & 0.007573 & 0.889247 & 0.009339 & 0.983760 & 0.001184 \\
USP6NL & 0.942667 & 0.004742 & 0.905583 & 0.009078 & 0.990921 & 0.000763 \\
HSPH1 & 0.944859 & 0.008508 & 0.909198 & 0.009486 & 0.985005 & 0.001432 \\
HSP90AB1 & 0.943538 & 0.005513 & 0.911158 & 0.009103 & 0.988477 & 0.001059 \\
PER1 & 0.974610 & 0.007156 & 0.900304 & 0.010561 & 0.981483 & 0.001727 \\
SPDYE3 & 0.937633 & 0.004811 & 0.879168 & 0.009912 & 0.976363 & 0.001818 \\
SERTAD1 & 0.931885 & 0.005830 & 0.899288 & 0.009403 & 0.981564 & 0.001535 \\
CASP7 & 0.953783 & 0.005448 & 0.905503 & 0.009068 & 0.988374 & 0.001003 \\
ABCA7 & 0.956684 & 0.003504 & 0.904476 & 0.008171 & 0.984454 & 0.001290 \\
LRIG1 & 0.966247 & 0.002927 & 0.907852 & 0.009604 & 0.972603 & 0.002650 \\
JAZF1 & 0.958753 & 0.004149 & 0.900445 & 0.009152 & 0.984720 & 0.001610 \\
BAG2 & 0.946831 & 0.005403 & 0.900583 & 0.010879 & 0.981685 & 0.001940 \\
SOD1 & 0.952839 & 0.005172 & 0.917126 & 0.009523 & 0.989726 & 0.001196 \\
BDNF & 0.926893 & 0.006588 & 0.900256 & 0.010109 & 0.980257 & 0.001792 \\
GRID2 & 0.914828 & 0.007292 & 0.915945 & 0.008164 & 0.988772 & 0.000990 \\
ZBTB11 & 0.969066 & 0.004345 & 0.890983 & 0.010178 & 0.979454 & 0.001782 \\
PRKD3 & 0.907688 & 0.009919 & 0.921829 & 0.009350 & 0.991708 & 0.001093 \\
RAB10 & 0.947483 & 0.005489 & 0.893835 & 0.009615 & 0.983929 & 0.001271 \\
TREM2 & 0.841656 & 0.009441 & 0.830463 & 0.010803 & 0.971587 & 0.002061 \\
IL1B & 0.878299 & 0.009188 & 0.883596 & 0.009095 & 0.982150 & 0.001461 \\
MAPT & 0.924418 & 0.009486 & 0.890104 & 0.010685 & 0.976285 & 0.002130 \\
TNFRSF1A & 0.928748 & 0.006825 & 0.886964 & 0.010375 & 0.978290 & 0.001829 \\
ZNF184 & 0.974879 & 0.003540 & 0.899847 & 0.010553 & 0.986731 & 0.001237 \\
HBP1 & 0.954659 & 0.003636 & 0.901980 & 0.008768 & 0.983792 & 0.001302 \\
SLC44A1 & 0.951556 & 0.003994 & 0.894284 & 0.009250 & 0.984850 & 0.001376 \\
SPI1 & 0.903774 & 0.007027 & 0.885691 & 0.008506 & 0.985816 & 0.000992 \\
SHARPIN & 0.950380 & 0.003898 & 0.904180 & 0.008871 & 0.986449 & 0.001108 \\
HS3ST5 & 0.945011 & 0.005517 & 0.925161 & 0.008172 & 0.986010 & 0.001408 \\
ZCWPW1 & 0.920636 & 0.006568 & 0.875920 & 0.010697 & 0.980991 & 0.001720 \\
FBXL7 & 0.948596 & 0.004417 & 0.894724 & 0.009339 & 0.978755 & 0.001654 \\
BHLHE40 & 0.956553 & 0.009276 & 0.929798 & 0.007867 & 0.989180 & 0.001133 \\
CLU & 0.954212 & 0.006760 & 0.919897 & 0.009964 & 0.984711 & 0.001906 \\
ADAMTS1 & 0.952468 & 0.003271 & 0.880563 & 0.009136 & 0.978367 & 0.001510 \\
SNX1 & 0.959517 & 0.003291 & 0.902419 & 0.009600 & 0.977847 & 0.001987 \\
BACE1 & 0.942698 & 0.005287 & 0.880767 & 0.009687 & 0.980354 & 0.001539 \\
SREBF1 & 0.944539 & 0.004576 & 0.895868 & 0.010175 & 0.980295 & 0.001683 \\
SORL1 & 0.970670 & 0.009098 & 0.909734 & 0.009874 & 0.981812 & 0.001985 \\
PLCG2 & 0.865604 & 0.009716 & 0.886153 & 0.009463 & 0.987293 & 0.000957 \\
PRNP & 0.967726 & 0.015970 & 0.915171 & 0.008849 & 0.989678 & 0.000979 \\
\bottomrule
\end{tabular}
\caption{Pearson Correlation (R) and Mean Squared Error (MSE) values, computed voxel-wise, comparing spatial gene expression between INR methods and \texttt{abagen}.}
\label{tab:r_mse}
\end{table}

\begin{table}[htbp]
\small
\ContinuedFloat
\centering
\setlength{\tabcolsep}{3pt}
\begin{tabular}{@{}l@{\hspace{2pt}}*{6}{c}@{}}
\toprule
& \multicolumn{2}{c}{INR} & \multicolumn{2}{c}{INR + Spectral Embedding} & \multicolumn{2}{c@{}}{%
\makecell{INR + Spectral Embedding + \\ Regional Averaged}}\\
\cmidrule(r){2-3}\cmidrule(r){4-5}\cmidrule{6-7}
Gene & R & MSE & R & MSE & R & MSE\\
\midrule
HSPA1L & 0.953559 & 0.006569 & 0.900637 & 0.009171 & 0.982243 & 0.001525 \\
DOC2A & 0.949602 & 0.012865 & 0.900651 & 0.009244 & 0.986003 & 0.001296 \\
PICALM & 0.943772 & 0.004393 & 0.895539 & 0.008760 & 0.977797 & 0.001776 \\
INPP5D & 0.954735 & 0.004857 & 0.905042 & 0.009644 & 0.981576 & 0.001667 \\
TYROBP & 0.956185 & 0.006919 & 0.903408 & 0.008689 & 0.985959 & 0.001117 \\
DNAJB4 & 0.963432 & 0.003432 & 0.910332 & 0.009293 & 0.984078 & 0.001488 \\
DDIT4 & 0.966846 & 0.005757 & 0.918492 & 0.010391 & 0.982549 & 0.002473 \\
SIRPA & 0.973182 & 0.013825 & 0.920415 & 0.008714 & 0.990002 & 0.000984 \\
TOMM40 & 0.916815 & 0.007326 & 0.912679 & 0.008849 & 0.987516 & 0.001124 \\
ATP8B1 & 0.959421 & 0.006598 & 0.907026 & 0.008640 & 0.991199 & 0.000721 \\
SORT1 & 0.931184 & 0.007358 & 0.910259 & 0.009106 & 0.985492 & 0.001416 \\
ADAM17 & 0.945668 & 0.004357 & 0.900326 & 0.008787 & 0.984191 & 0.001250 \\
TNIP1 & 0.952973 & 0.006711 & 0.903496 & 0.008947 & 0.983239 & 0.001479 \\
PTK2B & 0.931154 & 0.006899 & 0.902220 & 0.009580 & 0.985093 & 0.001284 \\
SPP1 & 0.946465 & 0.005260 & 0.897212 & 0.010231 & 0.980848 & 0.002301 \\
ABCA1 & 0.924276 & 0.005924 & 0.887454 & 0.010358 & 0.973353 & 0.002147 \\
EED & 0.981767 & 0.009480 & 0.904103 & 0.009367 & 0.983617 & 0.001411 \\
PILRA & 0.949643 & 0.005768 & 0.898947 & 0.010285 & 0.984018 & 0.001432 \\
COX7C & 0.951725 & 0.004492 & 0.898417 & 0.009816 & 0.979864 & 0.001778 \\
NYAP1 & 0.919314 & 0.006722 & 0.886625 & 0.009471 & 0.978889 & 0.001520 \\
PSEN2 & 0.948747 & 0.004419 & 0.903901 & 0.009614 & 0.986386 & 0.001196 \\
NEAT1 & 0.949688 & 0.004421 & 0.903814 & 0.009328 & 0.979013 & 0.001817 \\
APOE & 0.930564 & 0.006694 & 0.888540 & 0.011407 & 0.969402 & 0.002956 \\
TMEM41A & 0.923277 & 0.006671 & 0.896843 & 0.008845 & 0.984183 & 0.001417 \\
CX3CR1 & 0.953823 & 0.008054 & 0.910862 & 0.008390 & 0.986385 & 0.001140 \\
HS3ST2 & 0.968227 & 0.017745 & 0.900629 & 0.010388 & 0.975043 & 0.002320 \\
WDR12 & 0.947622 & 0.004912 & 0.902498 & 0.010033 & 0.981503 & 0.001696 \\
DNAJA1 & 0.980460 & 0.011175 & 0.921335 & 0.008152 & 0.990011 & 0.000934 \\
IER5 & 0.969479 & 0.015337 & 0.911407 & 0.009932 & 0.984620 & 0.001553 \\
PLD3 & 0.957528 & 0.012343 & 0.898418 & 0.010255 & 0.981399 & 0.001676 \\
ZBTB7A & 0.941094 & 0.005762 & 0.913327 & 0.009550 & 0.984982 & 0.001484 \\
NCK2 & 0.964858 & 0.006307 & 0.911694 & 0.009504 & 0.986564 & 0.001299 \\
GRIN2B & 0.926158 & 0.005888 & 0.883711 & 0.011352 & 0.981902 & 0.001512 \\
CREB3L4 & 0.953623 & 0.003800 & 0.886660 & 0.008967 & 0.985622 & 0.001043 \\
WDR81 & 0.963310 & 0.005167 & 0.900848 & 0.009117 & 0.982532 & 0.001455 \\
ANK3 & 0.914004 & 0.008443 & 0.909242 & 0.009523 & 0.984810 & 0.001453 \\
PVR & 0.902352 & 0.008763 & 0.893956 & 0.010139 & 0.982898 & 0.001526 \\
BIN1 & 0.967357 & 0.009394 & 0.883762 & 0.010613 & 0.979193 & 0.001715 \\
ANKH & 0.946471 & 0.007937 & 0.929455 & 0.007513 & 0.989556 & 0.001040 \\
EPDR1 & 0.916545 & 0.007420 & 0.902947 & 0.009991 & 0.987632 & 0.001130 \\
HSP90AA1 & 0.950659 & 0.005435 & 0.915146 & 0.008761 & 0.987747 & 0.001131 \\
APP & 0.939328 & 0.006284 & 0.914186 & 0.011159 & 0.982634 & 0.002121 \\
MINDY2 & 0.947459 & 0.004528 & 0.916849 & 0.009334 & 0.984879 & 0.001662 \\
HSPA8 & 0.964875 & 0.014243 & 0.903797 & 0.011438 & 0.986981 & 0.001362 \\
MAPK14 & 0.971500 & 0.010717 & 0.917127 & 0.009028 & 0.989449 & 0.001032 \\
REST & 0.947781 & 0.005753 & 0.898259 & 0.009293 & 0.982085 & 0.001461 \\
DNAJB6 & 0.981933 & 0.010880 & 0.899498 & 0.009667 & 0.978101 & 0.001951 \\
EGR1 & 0.961397 & 0.006763 & 0.920787 & 0.008768 & 0.987310 & 0.001273 \\
\bottomrule
\end{tabular}
\caption{Pearson Correlation (R) and Mean Squared Error (MSE) values, computed voxel-wise, comparing spatial gene expression between INR methods and \texttt{abagen}. (continued)} 
\label{tab:r_mse}
\end{table}

\begin{figure}[htbp]
    \centering
    \includegraphics[width=\textwidth]{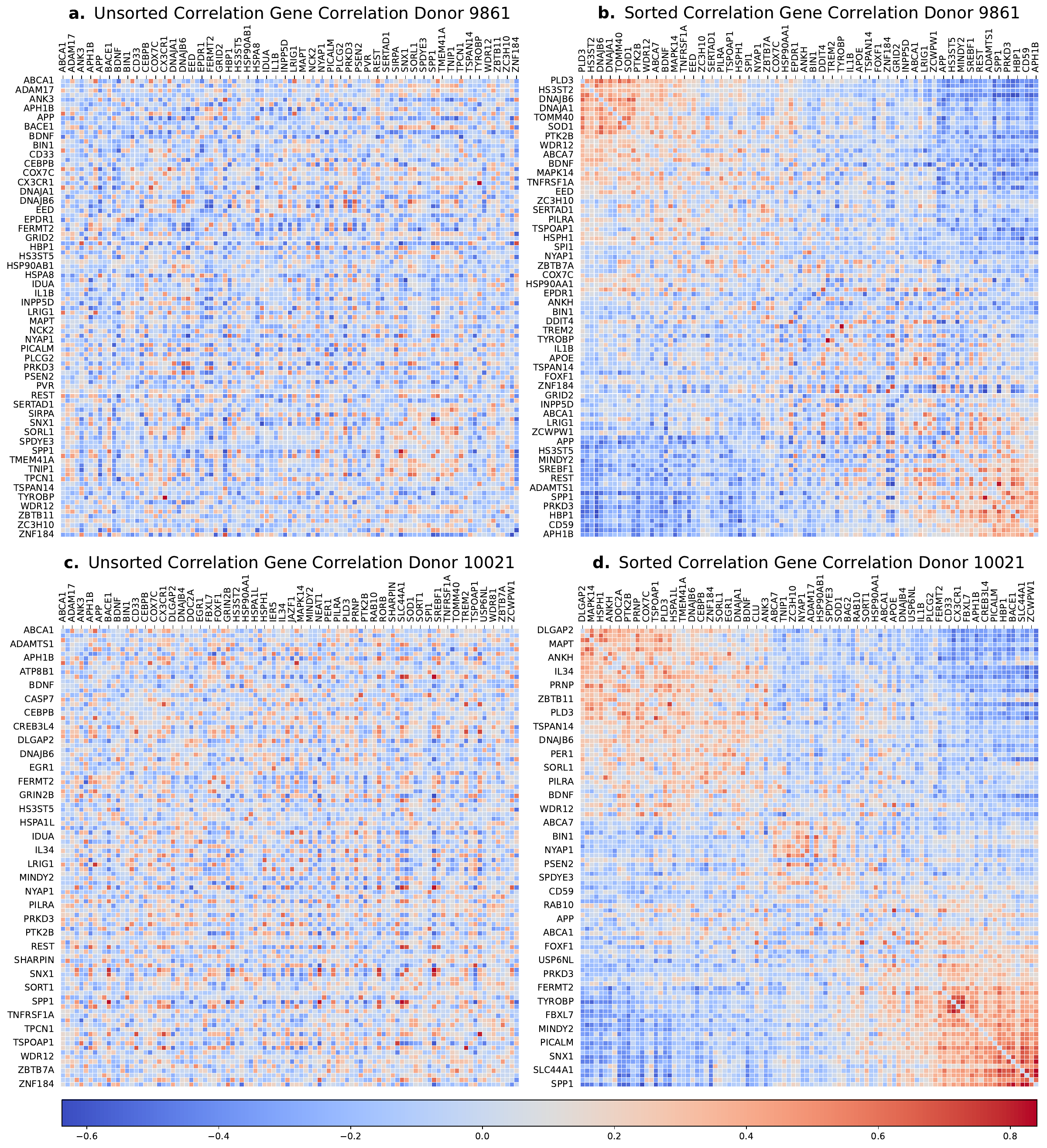}
    \caption{Side-by-side comparison of gene heatmaps, red implies positive correlations between genes and blue are negative correlations. (a) and (c) are original ordered genes alphabetically which has no obvious correlations for neighbor genes. (b) and (d) are sorted by spectral embedding orders therefore show positive correlations on nearby genes.}
    \label{fig:gene_heatmaps}
\end{figure}


\end{document}